\documentclass[twocolumn,showpacs,aps,pra,superscriptaddress]{revtex4}
\usepackage{graphicx}
\usepackage{dcolumn}
\usepackage{bm}
\usepackage{amsmath}
\usepackage{amssymb}
\usepackage{epsf}
\usepackage{longtable}
\newcommand{\be}{\begin{equation}}
\newcommand{\ee}{\end{equation}}
\newcommand{\bea}{\begin{eqnarray}}
\newcommand{\eea}{\end{eqnarray}}

\newcommand{\LD}{$\Lambda$-doublet~}
\newcommand{\neff}{n_{\footnotesize\mbox{eff}\footnotesize}}

\newcommand{\Neff}{N_{\footnotesize\mbox{eff}\footnotesize}}
\newcommand{\etaeff}{\eta_{\footnotesize\mbox{eff}\footnotesize}}

\begin{document}
\title{Prospects for the cavity-assisted laser cooling of molecules}

\author{Benjamin L. Lev}
\email{benlev@jila.colorado.edu}
\affiliation{JILA, National Institute of Standards and Technology and the University of Colorado \\ Department of Physics, University of Colorado, Boulder, Colorado 80309-0440, USA}
\author{Andr{\'{a}}s Vukics}
\affiliation{Institut f{\"{u}}r theoretische Physik, Universit{\"{a}}t Innsbruck, Technikerstr. 25, A-6020 Innsbruck, Austria}
\affiliation{Research Institute for Solid State Physics and Optics, P{.}O{.} Box 49, H-1525 Budapest, Hungary}
\author{Eric R. Hudson}
\altaffiliation{Present address:  Department of Physics, Yale University, New Haven, CT 06520 USA}
\author{ \ \ \ \ \ \ \ \ \ \ \ \ \ \ \ \ \ \ \ \ \ \ \ \ \ \ \ \ \ \ \ \ \ \ \ \ \ \ \ \ \ \ \ \ \ \ \ \ \ \ \ \ \ \ \ \ \ \ \ \ \ \ \ \ \ \ \ \ \ \ \ \ \ \ \ \ \ \ \ \ \ \ \ \ \ \ \ \ \ \ \ \ Brian C. Sawyer}
\affiliation{JILA, National Institute of Standards and Technology and the University of Colorado \\ Department of Physics, University of Colorado, Boulder, Colorado 80309-0440, USA}
\author{Peter Domokos}
\affiliation{Research Institute for Solid State Physics and Optics, P{.}O{.} Box 49, H-1525 Budapest, Hungary}
\author{Helmut Ritsch}
\affiliation{Institut f{\"{u}}r theoretische Physik, Universit{\"{a}}t Innsbruck, Technikerstr. 25, A-6020 Innsbruck, Austria}
\author{Jun Ye}
\affiliation{JILA, National Institute of Standards and Technology and the University of Colorado \\ Department of Physics, University of Colorado, Boulder, Colorado 80309-0440, USA}

\date{\today}
\begin{abstract} 

Cooling of molecules via free-space dissipative scattering of photons is thought not to be practicable due to the inherently large number of Raman loss channels available to molecules and the prohibitive expense of building multiple repumping laser systems.  The use of an optical cavity to enhance coherent Rayleigh scattering into a decaying cavity mode has been suggested as a potential method to mitigate Raman loss, thereby enabling the laser cooling of molecules to ultracold temperatures.  We discuss the possibility of cavity-assisted laser cooling particles without closed transitions, identify conditions necessary to achieve efficient cooling, and suggest solutions given experimental constraints.  Specifically, it is shown that cooperativities much greater than unity are required for cooling without loss, and that this could be achieved via the superradiant scattering associated with intracavity self-localization of the molecules.  Particular emphasis is given to the polar hydroxyl radical (OH), cold samples of which are readily obtained from Stark deceleration.

\end{abstract}
\maketitle

\section{Introduction}

The experimental realization of large samples of ultracold, ground state dipolar molecules would be a major breakthrough for research in fields as diverse as ultracold collisions and chemistry to quantum information processing and the study of novel correlated states of matter~\cite{Doyle04}.  In particular, the anisotropic dipole-dipole interaction becomes non-negligible for polar clouds below approximately 100 $\mu$K.  Exotic states of dipolar matter, such as field-linked states and dipolar crystals, may be observable in this regime~\cite{Bohn03}. If the rich field of ultracold alkali Feshbach physics is any measure, then ultracold molecular collisions and chemistry in the presence of the dipolar interaction and external electrical or magnetic fields promises to be fascinating~\cite{Bohn05,Krems05,Hudson06b}.  Moreover, the precision motional control attainable only at ultralow temperatures is crucial for constructing the architectures necessary to realize quantum logic gates or spin lattice simulations using the dipolar interactions~\cite{DeMille02,Zoller06}.

While many techniques for ultracold, ground state polar molecule production show promise, none so far have simultaneously yielded the low temperatures and high densities required to pursue these goals.  Photoassociation of ultracold atoms and subsequent optical pumping to the molecular ground state~\cite{DeMille05,Eyler04} has achieved lower sample temperatures ($\sim100$ $\mu$K) than techniques such as buffer gas cooling ($\sim400$ mK)~\cite{Doyle98}.  The Stark decelerator provides a nice compromise between density and temperature~\cite{Meijer99,Bochinski03}.  Electric and magnetic trapping of samples as cold as 10 mK at densities approaching $10^7$ cm$^{-3}$ have been demonstrated~\cite{Meijer05a,Sawyer07}.  However, new cooling techniques are required if we hope to push well below the 1 mK regime.

Unlike atoms, molecules typically have a large number of channels into which a given excited state can decay.  This makes the efficient free-space laser cooling~\cite{MetcalfBook99} of molecules challenging due to practical limits on the number of lasers one can build and operate to achieve ground state repumping after each photon scattering event~\cite{Gould96} (for a practical scheme involving only several repumpers, see~\cite{DiRosa04}).  Evaporative and sympathetic cooling techniques are quite promising, but require an initial density higher than what is currently available and is sensitive to the particular molecular species' collision cross-section, which is generally unknown~\cite{Bohn06}.  Cavity-assisted laser cooling~\cite{Vuletic00, Ritsch03_review} is a promising---though not fully understood---solution in that it provides dissipative cooling largely independent of the details of the molecular structure when the coupling between the molecules and the cavity is strong.  That the coupling needs to be strong is of prime importance for achieving the practical cooling of molecules.  Establishing this fact as well as exploring how to accomplish strong coupling between the cavity and molecular system is the main subject of this Article.
\begin{figure}[t]
\begin{center}
\scalebox{0.32}[0.32]{\includegraphics{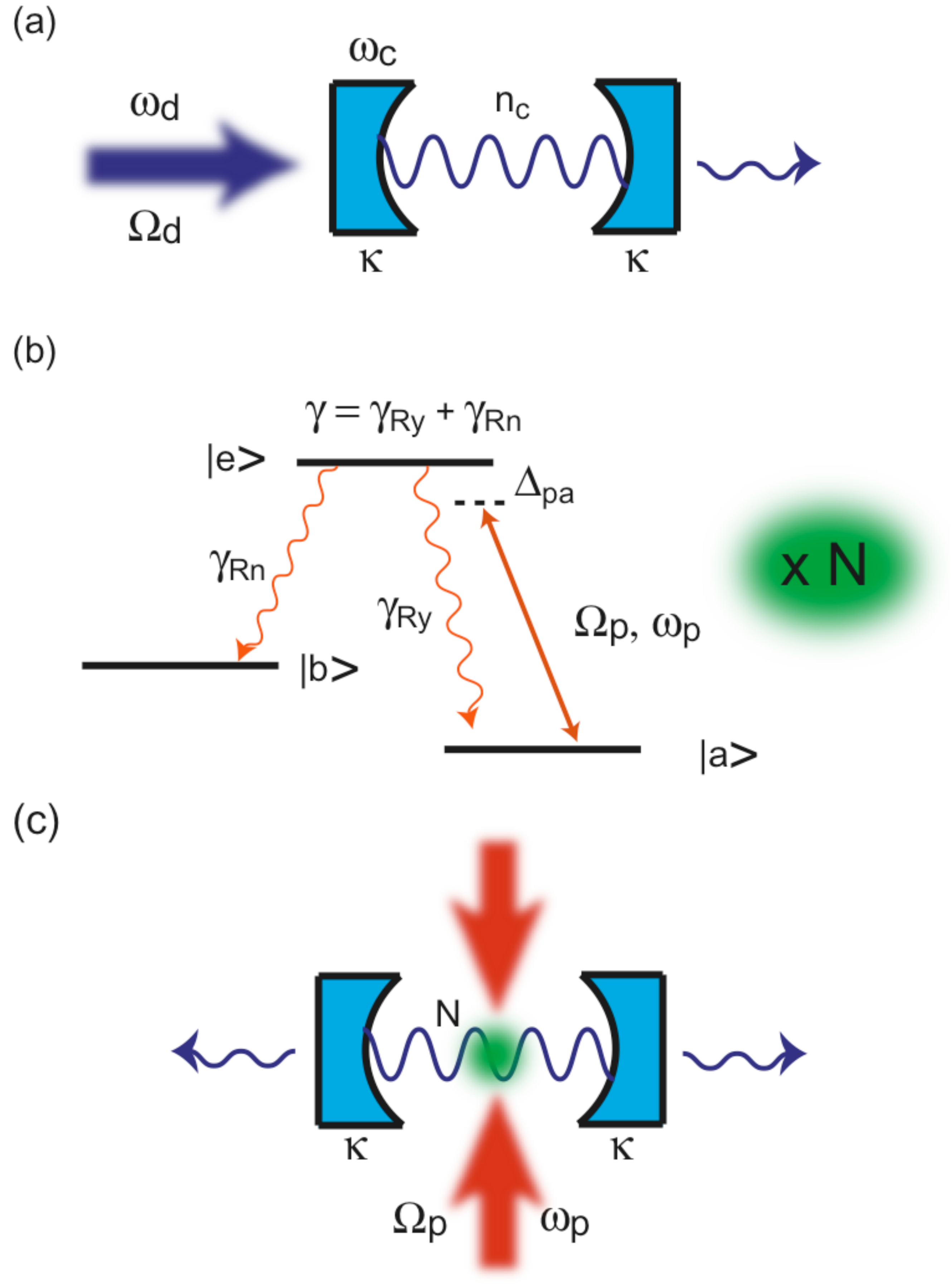}}
\caption{\label{fig:scenario}a) Fabry-Perot cavity with resonance frequency $\omega_c$ driven by a laser of frequency $\omega_d$ and drive strength $\Omega_d$, defined in Eq{.}~\ref{nc}.  In steady-state, there are $n_c=\langle a^\dagger a\rangle$ intracavity photons which escape the cavity via the mirrors at rate $2\kappa$. b) Three-level representation of the $N$ intracavity molecules.  A classical laser field of Rabi frequency, $\Omega_p$, and frequency, $\omega_p$, couples the ground state $|a\rangle$ to the electronically excited state $|e\rangle$, whose frequency difference is $\omega_a$.  Excitations decay at the total rate $\gamma$.  Assuming $|e\rangle$ is unsaturated, population is elastically (Rayleigh) scattered back to $|a\rangle$ at the rate $\gamma_{Ry}$. Population is lost to the myriad molecular states, represented collectively by $|b\rangle$, at the Raman scattering rate $\gamma_{Rn}$.  The difference between $\omega_a$ and the transition frequency, $\omega_b$, between $|b\rangle$ and $|e\rangle$ is much larger than $\Delta_{pa}$.  c) The molecules-cavity system may be pumped by the cavity drive laser, a transverse pump beam (shown), or both.  The transverse pump beam of frequency $\omega_p$ and Rabi frequency, $\Omega_p$, is typically red-detuned from both the cavity and the molecular resonance ($\omega_p<\omega_c\ll\omega_a$). The $N$ molecules may be trapped or transiently passing through the cavity mode.}
\end{center}
\end{figure}
\begin{figure}[t]
\begin{center}
\scalebox{0.5}[0.5]{\includegraphics{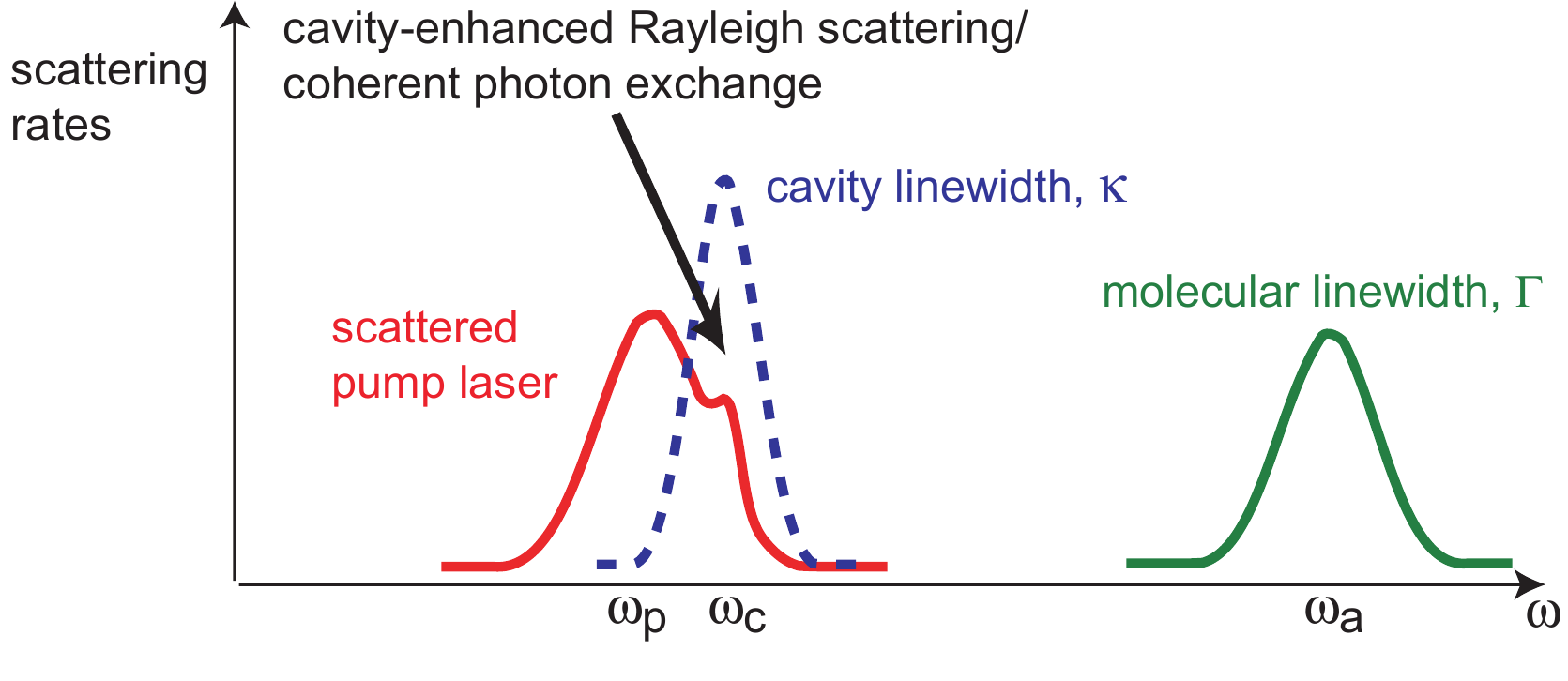}}
\caption{\label{fig:energy}The transverse pump laser (red) scatters coherently from the far-detuned atomic medium (green).  This laser does not excite much population, but its energy distribution can be Doppler broadened due to the velocity spread of the molecular cloud.  The blue-detuned and overlapping cavity resonance (dashed) modifies the scattered spectrum to favor higher energy photons, thus extracting energy from the atomic motion.  Figure adapted from Ref{.}~\cite{Vuletic05}.}
\end{center}
\end{figure} 

Cavity-assisted laser cooling of molecules brings together many otherwise disparate disciplines:  cavity quantum electrodynamics (QED), molecular physics, and laser cooling and trapping.  In this Article, we describe the physics of cavity-cooling species that do not possess closed transitions, and in doing so, we review some elements of cavity QED and atom-photon interactions so as to build a common language amenable to researchers from all three fields.  We begin by reviewing the cavity cooling of a two-level system.  This is followed by a discussion of molecules in cavities.  Whereas strong coupling is not necessary for cooling two-level systems, it is shown that it is crucial for cooling molecular samples.  We conclude by analyzing methods for increasing the efficiency of cooling molecules given the constraints posed by experimentally realizable cavity geometry, cavity quality, and sample densities and temperatures.  For a concrete discussion, we focus our attention on the OH radical, a highly polarizable molecule well suited for exploring dipolar physics~\cite{Bohn02,Bohn03}, cold collisions~\cite{Hudson06b,GerardOHXe06},  precision measurements of fundamental constants~\cite{Hudson06,Lev06}, and quantum information processing~\cite{DeMille02,Lev06}.

The cavity cooling system is illustrated in Fig{.}~\ref{fig:scenario}, in which a standing wave cavity is combined with a three-level system.  While cavity cooling was originally discussed in the context of cooling via driving the cavity mode~\cite{Ritsch97, Vuletic00}, we will focus on pumping via transverse beams, which provides strong cooling in all three dimensions~\cite{Vuletic01,Ritsch02_trans}.  Figure~\ref{fig:scenario}(c) depicts the configuration for 3-D transverse pumping, though the two pump beams in the plane perpendicular to the page are not shown.  Throughout this Article, unless otherwise stated, discussion is confined to the transverse pumping scenario.  

From a semiclassical viewpoint, the molecules act as a nonlinear element that coherently transfer excitation from the red-detuned transverse pump field to higher-frequency (bluer) intracavity field.  Because the blue-detuned field leaks out of the cavity stochastically, energy proportional to the fields' frequency difference is extracted from the system at the rate $2\kappa$.  Consequently, the molecules' motion is cooled, as this is the only mechanism from which to extract this energy difference.  The energetics of this cooling mechanism are shown in Fig{.}~\ref{fig:energy}.  

The primary purpose of this Article is to demonstrate that to make this molecular cavity cooling mechanism efficient---and therefore practical---one more condition must be satisfied.  Namely, the molecules must coherently scatter photons into the cavity mode at a rate much faster than the scattering into free-space.  This is important not because the free-space scattering leads to additional recoil heating, but because free-space scattering allows the molecules to relax back to the multitude of metastable molecular states---spin-orbit, rotational, and vibrational---via inelastic, Raman scattering.  This prematurely quenches the cooling process because these states are effectively dark to the very far off-resonant transverse pump laser and will not be repumped.  We address the conditions that must be satisfied to ensure that this Raman loss rate is less than the cooling rate, and discuss experimental constraints to these conditions.

\section{Single particle master equation analysis}

The previous section described the general idea of cavity-assisted laser cooling, but greater insight into the problem is required to evaluate its effectiveness for cooling molecules with realistic internal energy level distributions.  Solutions to the master equation for the joint molecule-cavity system allow the identification of optimal experimental conditions for achieving cavity-assisted laser cooling.  However, to gain intuition, we first study the case of a single, motionless, two-level atom coupled to only one cavity mode.  In the following sections, we use the term ``atom" when discussing general cavity QED and cavity cooling, and reserve ``molecule" to when we discuss phenomena or conditions peculiar to cooling multi-level systems.

The derivation of the interaction-picture Hamiltonian begins by applying the unitary transformation $U=\exp\left[-i\omega_pt(\hat{a}^\dagger\hat{a}+\hat{\sigma}_+\hat{\sigma}_-)\right]$ to the joint atom-cavity system Hamiltonian, which assumes that the drive and pump frequencies are approximately equal ($\omega_d\approx\omega_p$). The electric dipole and rotating wave approximations are then applied to arrive at the Hamiltonian describing the coherent dynamics in the presence of both the drive and the pump fields: 
\bea\label{H_1atom1mode} \hat{H}&=&-\Delta_{pa}\hat{\sigma}_+\hat{\sigma}_--\Delta_{pc}\hat{a}^\dagger\hat{a}+g\left[\hat{a}^\dagger\hat{\sigma}_-+\hat{\sigma}_+\hat{a}\right]\nonumber\\
& &+\Omega_p[\hat{\sigma}_-+\hat{\sigma}_+]/2+\Omega_d[\hat{a}+\hat{a}^\dagger]/2.
\eea
In this equation, $\hat{\sigma}_-$ is the atomic lowering operator, and $\hat{a}$ is the cavity field annihilation operator.  The first two terms are the bare atom and cavity energies, with $\Delta_{pa}\equiv\omega_p-\omega_a$ and $\Delta_{pc}\equiv\omega_p-\omega_c$, while the atomic pump and cavity drive terms are the fourth and fifth, respectively.  The atom-cavity detuning is $\Delta_{ca}\equiv\Delta_{pc}-\Delta_{pa}=\omega_a-\omega_c$.  

The third term represents the atom-cavity interaction:  excitation is coherently exchanged at the rate $g$, which depends on the atom's position in the intracavity mode structure and is inversely proportional to the square-root of the cavity mode volume.  Specifically, $g=g_0\psi(\hat{r})$ with \be  \hbar g=\vec{\mu}\cdot\mathcal{\vec{E}}=\psi(\hat{r})\mu\sqrt{\frac{\hbar\omega}{2 e_0 V_m}}=\psi(\hat{r})\hbar\sqrt{\frac{3c\lambda^2\gamma_{\perp}}{4\pi V_m}},\ee where $\psi(\hat{r})\leq1$ accounts for the atom's position, $\vec{\mu}$ is the transition dipole moment, and $\mathcal{\vec{E}}$ is the electric field.   We have explicitly included Plank's constant in the above equation for clarity, but will set $\hbar=1$ in all subsequent equations, including Eq.~\ref{H_1atom1mode}.  The mode volume is $V_m\approx\pi w_0^2 L/4$, where $w_0$ and $L$ are the cavity waist and length, respectively.  In the limit that the atomic linewidth is dominated solely by radiative processes, $\gamma=2\gamma_{\perp}$, where $\gamma_{\perp}$ is the decay rate of the atomic dipole.  In a sense, $2 g$ is the Rabi flopping rate of the atom stimulated by the vacuum field of the cavity.  The purpose of making small cavities in single-atom cavity QED research~\cite{BermanBook94, Mabuchi02} is to enhance the electric field associated with a single photon so that this zero-point field fluctuation has a non-negligible effect on the atom.  

When one has an atom in a transversely pumped cavity, the situation is not unlike that of a free space atom interrogated by two lasers with Rabi rates $\Omega_p$ and $2g\sqrt{n_c}$, where $n_c=\langle\hat{a}^\dagger\hat{a}\rangle_0$ is the number of intracavity photons in the absence of an atom.  The presence of the atom in the cavity modifies the number of intracavity photons, and this back reaction of the atom onto the intracavity field produces a self-consistent field intensity that is not necessarily equal to $\langle\hat{a}^\dagger\hat{a}\rangle_0$.

The master equation for the density matrix, $\hat{\rho}$, of the joint state of the atom and cavity with a transverse pump field is the following: 
\bea\label{2levelmaster}
\dot{\hat{\rho}}&=&-i\left[\hat{H},\hat{\rho}\right]+\gamma_{\perp}(2\hat{\sigma}_-\hat{\rho}\hat{\sigma}_+-\hat{\sigma}_+\hat{\sigma}_-\hat{\rho}-\hat{\rho}\hat{\sigma}_+\hat{\sigma}_-) \nonumber \\
& &+\kappa(2\hat{a}\hat{\rho}\hat{a}^\dagger-\hat{a}^\dagger\hat{a}\hat{\rho}-\hat{\rho}\hat{a}^\dagger\hat{a}).
\eea
The cavity energy decay rate is $2\kappa$, and $n_c$ is equal to \be\label{nc} n_c=\frac{\Omega^2_d/4}{\kappa^2+\Delta^2_{pc}}.\ee  All optical fields are far detuned from the atomic resonance in the cavity cooling scenario, and typically $\Delta_{pc}\ll[\Delta_{pa},\Delta_{ca}]$.

\begin{figure}[t]
\begin{center}
\scalebox{0.5}[0.5]{\includegraphics{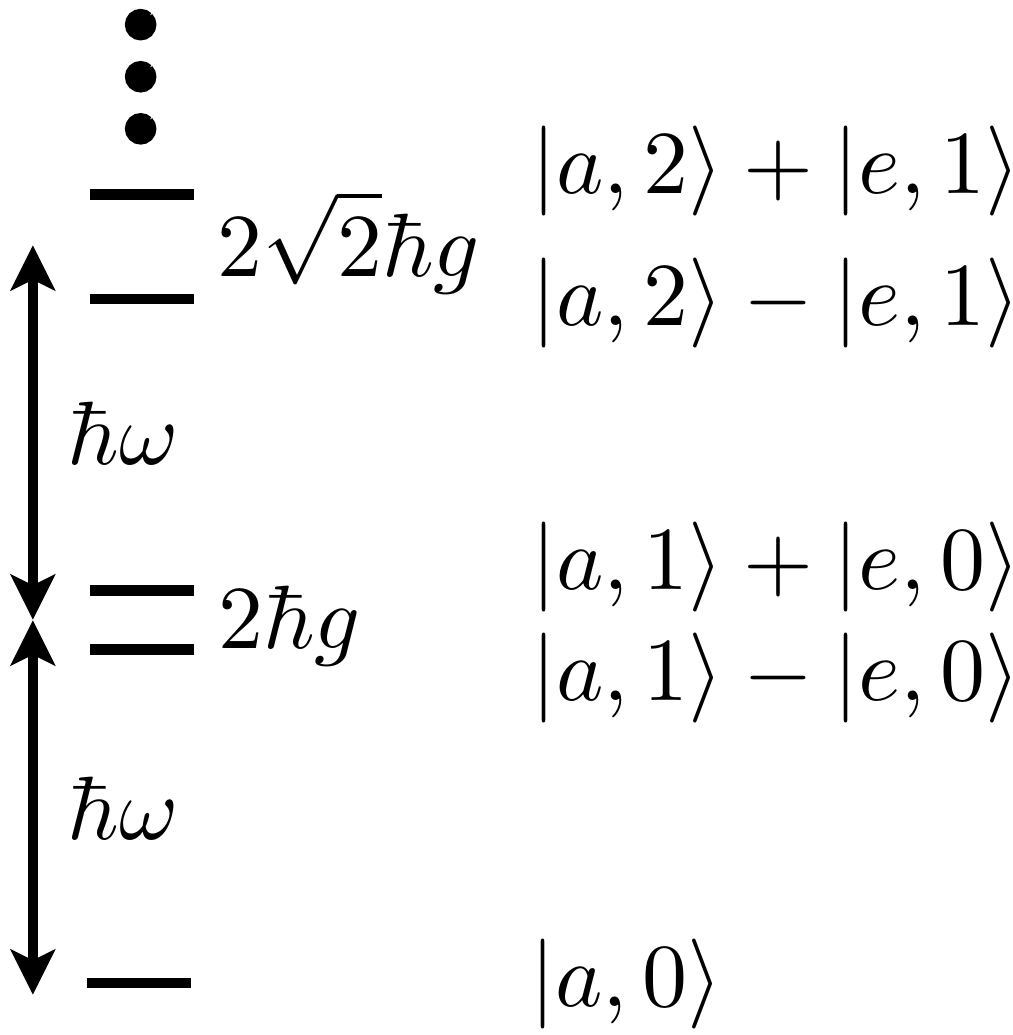}}
\caption{\label{fig:JC}Jaynes-Cummings ladder representing the spectrum of joint atom-cavity eigenstates for the case when the atom and cavity are on resonance ($\omega_c=\omega_a$).  The kets $|\mbox{atom},\mbox{cavity}\rangle$ denote the joint states.  The ground state, $|a,0\rangle=|g\rangle$, has zero excitation while the first band of excited states are symmetric and antisymmetric superpositions of one excitation in either the atomic $|e,0\rangle$ or cavity $|a,1\rangle$ modes.  These so-called dressed states~\cite{CohenTannoudji} are coupled by the ``single-photon" Rabi rate, $2g$.  The next excited band is similar, but with a $\sqrt{n_c}$ enhancement of the coupling~\cite{BermanBook94}.}
\end{center}
\end{figure}

Steady-state analytic solutions to the master equation may be obtained in the limit that coherences between the atom and cavity vanish on time scales much faster than any other in the problem.  The resulting semiclassical equations do not properly account for quantum fluctuations of joint atom-cavity excitations.  Nevertheless,  it may be shown that it reliably models the single atom-cavity behavior in the limit of: 1) weak driving wherein the atom is unsaturated---a condition we will later adopt;  2) very strong driving where the atom is completely saturated; and 3) when there are several intracavity atoms.  
Cases 1 \& 2 may be understood from the spectrum of joint eigenstates for the atom-cavity system shown in Fig{.}~\ref{fig:JC}.  Weak driving means that the first excited ``rung" on the ladder is minimally populated, while in strong driving, many rungs are populated and the excitation spacing becomes approximately equal, i.e., $\sqrt{n_c}\approx\sqrt{n_c+1}$.  In both cases, the dynamics may be treated semiclassically.

We will discuss case 3 separately in Section~\ref{multiatom}, but essentially, the uncorrelated nature of the atoms averages away the quantum fluctuations, even in the intermediate regime between weak and strong driving.  The multiatom case exhibits optical bistability in accordance to the solutions admitted by the nonlinear semiclassical (optical-Bloch) equations.  In contrast, a bilinear (master) equation describes the coupling of a single atom to the cavity mode, and thus, does not exhibit nonlinear effects in steady-state even though the atom-cavity coupling may be strong.  This is due to the dominance of quantum fluctuations in the single atom case that, in a sense, ``washes-out" the nonlinear effects~\cite{HoodThesis}.  We mention this to highlight the fact that in the far-detuned, many particle regime of the cavity-cooling scheme, semiclassical approximations are appropriate and will be used throughout this Article.

To obtain the steady-state semiclassical solutions, we make the approximation that all mixed operator expectations are factorable, e.g., $\langle\hat{a}^\dagger\hat{\sigma}_-\rangle\approx\langle\hat{a}^\dagger\rangle\langle\hat{\sigma}_-\rangle$.  This allows us to separate Eq{.}~\ref{2levelmaster} into two Hamiltonians:  one for the atom, $\hat{H}_a$, in which the cavity operator is converted to a C-number $\hat{a}\rightarrow\alpha$; and one for the cavity, $\hat{H}_c$, with $\hat{\sigma}_-\rightarrow\zeta$, where $\alpha$ is the field amplitude and $\zeta$ is the atomic dipole:  \bea 
\hat{H}_a&=&-\Delta_{pa}\hat{\sigma}_+\hat{\sigma}_-+g\left[\alpha^*\hat{\sigma}_-+\alpha\hat{\sigma}_+\right]+\Omega_p[\hat{\sigma}_++\hat{\sigma}_-]/2 \nonumber\\ &=&-\Delta_{pa}\hat{\sigma}_+\hat{\sigma}_-+\Omega'_p[\hat{\sigma}_++\hat{\sigma}_-]/2 \\
\hat{H}_c&=&-\Delta_{pc}\hat{a}^\dagger\hat{a}+g\left[\zeta\hat{a}^\dagger+\zeta^*\hat{a}\right]+\Omega_d[\hat{a}+\hat{a}^\dagger]/2 \nonumber\\ &=&-\Delta_{pc}\hat{a}^\dagger\hat{a}+\Omega'_d[\hat{a}+\hat{a}^\dagger]/2.
\eea
The second lines use the effective atom drive $\Omega'_p=2g\alpha+\Omega_p$ and cavity drive $\Omega'_d=2g\zeta+\Omega_d$ for clarity.  The master equation may be similarly separated:
\bea\label{2levelmaster_sep}
\dot{\rho_a}&=&-i\left[\hat{H_a},\rho_a\right]+\gamma_{\perp}(2\hat{\sigma}_-\rho_a\hat{\sigma}_+-\hat{\sigma}_+\hat{\sigma}_-\rho_a-\rho_a\hat{\sigma}_+\hat{\sigma}_-) \nonumber \\
\dot{\rho_c}&=&-i\left[\hat{H_c},\rho_c\right]+\kappa(2\hat{a}\rho_c\hat{a}^\dagger-\hat{a}^\dagger\hat{a}\rho_c-\rho_c\hat{a}^\dagger\hat{a}).
\eea

Setting $\dot{\rho}_a=0=\dot{\rho}_c$,  we obtain for the steady-state cavity field amplitude:  \bea\label{alpha}
\alpha=\mbox{Tr}[\rho_c\hat{a}]=\frac{\Omega'_d/2}{\Delta_{pc}+i\kappa}, 
\eea 
and for the atomic coherence: \bea\label{beta}
\zeta=\mbox{Tr}[\rho_a\hat{\sigma}_-]=\frac{\Omega'_p/2(\Delta_{pa}-i\gamma_{\perp})}{|\Omega'_p|^2/2+\gamma_{\perp}^2+\Delta^2_{pa}},
\eea
from which we obtain the atomic excited state population: \bea
\sigma_{ee}=\frac{|\Omega'_p/2|^2}{|\Omega'_p|^2/2+\gamma_{\perp}^2+\Delta^2_{pa}}.
\eea
To solve the system of Eqs{.}~\ref{alpha} and~\ref{beta} in closed-form, one needs to assume that the atomic excited state is not saturated, i.e., $\gamma_{\perp}^2+\Delta^2_{pa}\gg\Omega'^2_p$.  Fortunately, this is exactly the same weak driving condition we must satisfy to prevent the atom or molecule from incoherently scattering.  

The saturation parameter, $s$, characterizes the boundary between incoherent and coherent scattering:  \be s=\frac{\Omega_p^2/2}{\Delta_{pa}^2+\gamma^2/4}, \ee defined here in the absence of a cavity field.  When below $s\sim0.01$, the frequency of coherently scattered photons matches the incoming, and they are emitted in a dipole pattern with respect to the drive beam.  In contrast, a saturated atom scatters photons incoherently into $4\pi$ with an atomic linewidth broadened spectrum.  The rate of scattered photons is suppressed in the unsaturated regime, as may be seen from the following relation:   \be \Gamma_a=\frac{\gamma}{2}\frac{s}{(1+s)}=\gamma\sigma_{ee}. \ee In the limit that $s\leq1$, \be \zeta\approx\frac{\Omega'_p/2}{\Delta_{pa}+i\gamma_{\perp}}, \ee  and the cavity field amplitude becomes: \be \alpha\approx\frac{g\Omega_p/2+\Omega_d(\Delta_{pa}+i\gamma_{\perp})/2}{(\Delta_{pa}+i\gamma_{\perp})(\Delta_{pc}+i\kappa)-g^2}.\ee  

\section{The role of cooperativity}

For the discussion in this section, we will set the cavity drive to zero, $\Omega_d=0$, and discuss cavity cooling via transverse pumping solely, as sketched in Fig{.}~\ref{fig:scenario}(c).  We will address $\Omega_d\neq0$ in a subsequent section.  Energy enters the system only through the pump beam, whose intensity is proportional to $\Omega_p^2$.  Energy leaves the system either via the cavity mirrors at a rate proportional to the intracavity field, $\Gamma_c=2\kappa|\alpha|^2$, or by relaxation of the atomic excited state to modes other than the cavity, at rate $\Gamma_a$.  We assume that the solid angle subtended by the cavity mirrors is much less than $4\pi$.  Indeed, for all cavities of interest, the fractional solid angle is $\sim10^{-5}$, and $\Gamma_a$ is effectively unmodified by the cavity's presence.  In light of the discussion in the Introduction, $\Gamma_c$ is proportional to the cooling rate and therefore desirable to enhance, whereas the photons scattered at the rate $\Gamma_a$ lead to recoil heating and possibly to inelastic transitions---as in the case of molecular systems---and is therefore to be minimized.  

To ensure low atomic saturation, we now apply the dispersive weak driving condition, \be\label{conditions} \Delta_{pa}\gg\left[\Omega_{pa},\gamma_{\perp},g\right],\ee to arrive at the following rate expressions: \bea\label{cavscat1} \Gamma_c&\approx&2\kappa\frac{\Omega_p^2}{4\Delta_{pa}^2}\frac{g^2}{\Delta_{pc}^2+\kappa^2}, \\ \label{atomscat}\Gamma_a&\approx&2\gamma_{\perp}\frac{\Omega_p^2}{4\Delta_{pa}^2}.\eea  In the previous expressions, we have made the additional assumption that $\Delta_{pa}\gg\Delta_{pc}$, which is true for optimal cooling, which in turn requires $\Delta_{pc}\approx-\kappa$.  In light of the previous statement, Eq{.}~\ref{cavscat1} equals\be\label{cavscat2}  \Gamma_c\approx2\kappa\frac{\Omega_p^2}{8\Delta_{pa}^2}\frac{g^2}{\kappa^2},\ee and the ratio of scattered photons that contribute to cooling to those contributing to heating (and Raman loss in the case of molecules) is: \be\label{coop} C=\frac{\Gamma_c}{\Gamma_a}=\frac{g^2}{2\kappa\gamma_{\perp}}.\ee 

The label ``$C$" is chosen to highlight that this enhancement factor is none other than the single-atom cooperativity parameter well-known from cavity QED~\cite{BermanBook94}.  In essence, the condition $C\gg1$ signifies that the coherent atom-cavity dynamics dominate over dissipative.  In other words, the cavity field is stimulating the transfer of photons from the pump field to the decaying cavity mode at a rate faster than free-space scattering.  The cooperativity may be better understood by noting that the decay rate of the coupled atom-cavity system, in this limit, is $\Gamma_{ac}\approx\gamma(1+2C)\Omega^2_p/\Delta^2_{pa}$~\cite{BermanBook94}.  The extra factor $2C$ comes from the increased rate at which the pump field is depleted.  Important to note is the independence of the cavity-to-free space scattering ratio, $C$, on detuning.  Contrary to expectations, one cannot escape the free-space scattering that produces Raman loss by detuning far from resonance.  Detuning only leads to slower cooling, since both rates depend on the saturation parameter in the same fashion:  once incoherent scattering is quenched, there is no further advantage to increased detuning or decreasing pump intensity.  

\section{Cavity cooling in the perturbative limit}\label{perturb}

When $C<1$ (weak-coupling limit, assuming $g<\left[\kappa,\gamma_{\perp}\right]$), the cooling process may be described perturbatively as coherent Rayleigh scattering, as was done by {Vuleti\'{c}} {\it{et al.}}~\cite{Vuletic00,Vuletic01}.  A remark on language:  ``coherent" refers to scattering in the $s\leq1$ regime, wherein a definite relationship exists between the scattered field and the oscillating atomic dipole. ``Rayleigh" refers to the fact that any light scattered into the cavity mode is of a frequency such that the atom relaxes to the ground state of origin, and is thus an elastic process.  This occurs when $\Delta_{pc}$ is much smaller than any frequency difference, $\omega_{ab}$, between $|a\rangle$ and metastable ground states, $|b\rangle$.  The phrase ``coherent Rayleigh" means that the Rayleigh scattering is in a dipole pattern and is of the same frequency and spectral bandwidth as the pump field.  

For consistency, we remark that while the notion of coherence in Raman scattering---which is inherently an inelastic process---is not commonly defined, we take the phrase ``coherent Raman scattering" to mean that the character of the scattered field retains a definite relationship to the pump field, which is the case in the $s\ll1$ regime.  While there is a frequency offset between the pump and scattered field, there is a fixed phase relationship between them, and the field is scattered in a dipole pattern.  As will be discussed extensively in Section~\ref{molecule}, this coherent Raman scattering, which quenches the molecular cooling process, cannot be overcome by making $s$ small. The only remedy is to enhance the molecule-cavity coupling.
 
Since $g<\kappa$ in this limit, ``scattering" is the term of choice because any photon emitted into the cavity escapes via the mirrors before being reabsorbed by the atom.  Thus, the cavity field is incidental: the atom scatters photons from the incoming pump beam to the two beams emanating from the cavity mirrors and the cavity itself serves only to provide a concentrated density of states, which modifies the frequency spectrum of scattered photons.  The language of the Purcell effect---enhanced emission into the solid angle subtended by the cavity mirrors---is apt in this situation and {Vuleti\'{c}} {\it{et al.}} makes use of it to describe these dynamics.  As shown in Appendix~\ref{cavityparameters}, the Purcell factor is exactly equivalent to the cooperativity, but expressed in a more experimentally recognizable form.  We will return to this formalism when discussing multimode cavities in Section~\ref{multimode}.

The cooling force and rate can be readily derived in the perturbative regime, as was done by {Vuleti\'{c}} {\it{et al.}} for the case of cavity pumping ($\Omega_d\neq0$, $\Omega_p=0$)~\cite{Vuletic00} and transverse pumping ($\Omega_d=0$, $\Omega_p\neq0$)~\cite{Vuletic01}.  This is done without invoking the approximation $kv\leq\kappa$, where $k$ is the optical transition wavenumber and $v$ is the atomic velocity.   We do not revisit these derivations, but merely describe the conditions for obtaining optimal cooling rates and forces for the transverse pumping case.

\begin{figure}[t]
\begin{center}
\scalebox{0.5}[0.5]{\includegraphics{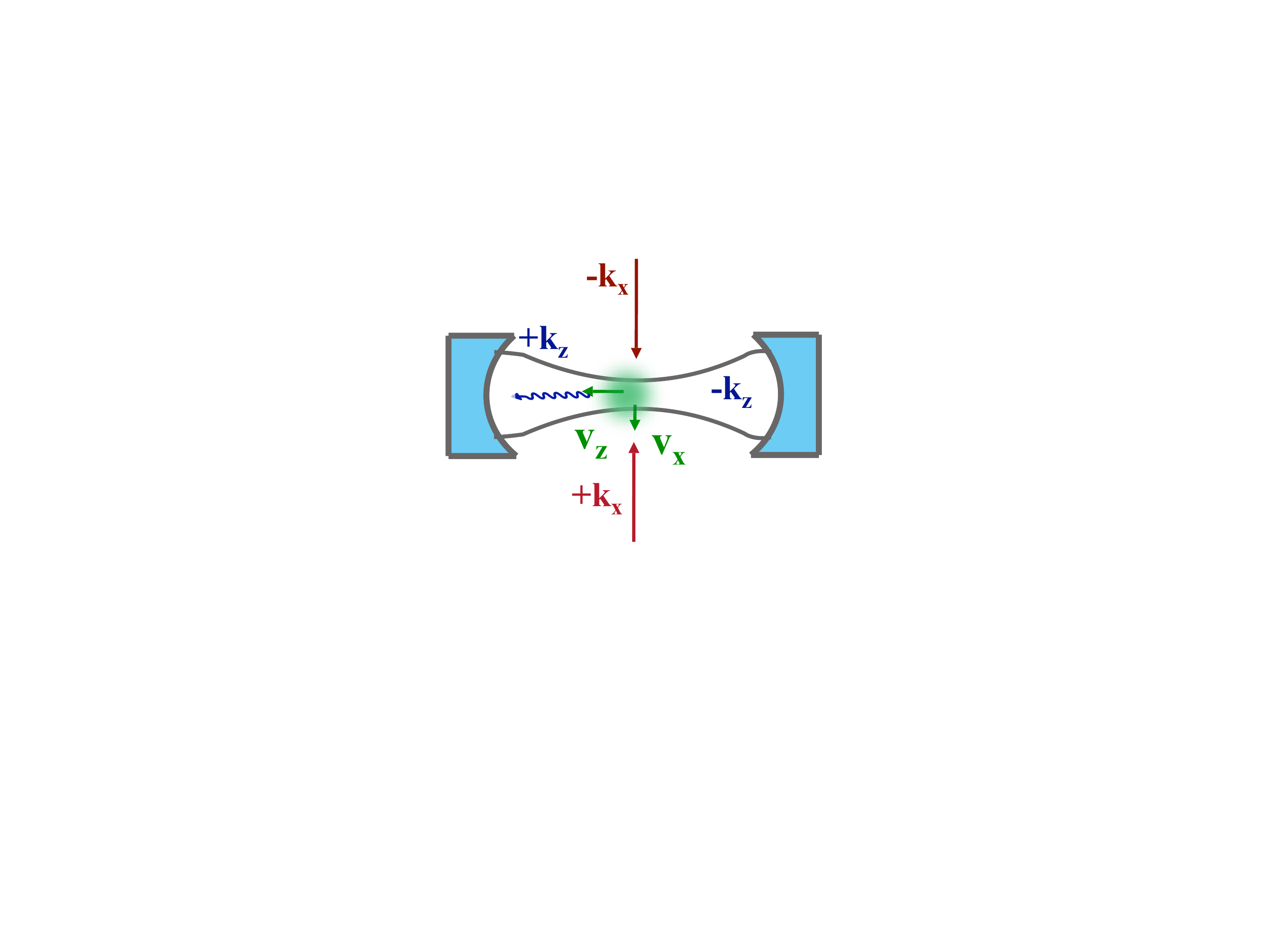}}
\caption{\label{fig:force}Illustration of the cavity cooling mechanism present in the perturbative (weak coupling) limit.  This mechanism can be referred to as cavity Doppler cooling~\cite{Vuletic00} in analogy to free-space Doppler cooling.  A scatterer preferentially absorbs red-detuned photons from the $+k_x$ beam.  Its forward velocity in $+k_z$ blue shifts emitted photons into resonance with the blue-detuned cavity, enhancing the scattering rate in this direction.  As a consequence of this two-photon process, the scatterer receives a momentum kick in a direction opposite to its combined $v_z$ and $v_x$ velocities and is therefore cooled.  A similar process works for cooling in $\hat{y}$ with the addition of another pump laser pair. Figure adapted from Ref{.}~\cite{Vuletic01}.}
\end{center}
\end{figure}

In the coherent Rayleigh scattering process, the energy extracted is equal to the difference in energy between the cavity resonance and the frequency of the pump beam.  The rate of cooling and damping force, however, depend on the spectral overlap between the scattered pump field and the cavity resonance.  Figure~\ref{fig:force} shows the coordinate system for cooling in the 2D plane spanned by one set of transverse pump beams and the cavity axis.  Cooling occurs via a two-photon scattering process in this perturbative limit, with a lower energy photon absorbed along $\pm \mathbf{k}_x$ and a higher energy one emitted along $\pm \mathbf{k}_z$.  The cooling force is thus along the diagonals spanned by $\mathbf{k}_x$ and $\mathbf{k}_z$.  

The scattering rate into the cavity from a single transverse beam(e{.}g{.}, $+\mathbf{k}_x$)  may be found by incorporating the Doppler shift $ \mathbf{k}\cdot\mathbf{v}$ into the $\Delta_{pc}$ detuning parameter of Eq{.}~\ref{cavscat1}.  Note that we assume $|\Delta_{pa}|\gg |kv|$.  Making the substitution $\Delta_{pc}\rightarrow|\Delta_{pc}|+ \mathbf{k}_x\cdot\mathbf{v}$, we have:  \be\label{dopplercooling} \Gamma_c(\mathbf{v})=2\kappa\frac{\Omega_p^2}{4\Delta_{pa}^2}\frac{g^2}{(|\Delta_{pc}|+ \mathbf{k}_x\cdot\mathbf{v})^2+\kappa^2}.\ee  In the above expression, we have ignored the recoil-induced detuning, $\omega_{rec}=\hbar k^2/2m$, since this is 100 times smaller than the Doppler shift for the $\sim$10 m/s OH samples we consider here.  The cooling power from this beam is: \be P_{cooling}\propto\hbar kv\Gamma_c(\mathbf{v}),\ee and the cooling rate for a velocity, $v$, that is much larger than the recoil velocity $\hbar k/m$, is: \be \Gamma_{cooling}=\frac{P_{cooling}}{E}\propto\frac{\hbar kv\Gamma_c(\mathbf{v})}{mv^2/2},\ee where $E$ is the atom's kinetic energy. From this equation, we see that the cooling rate is optimal when the pump detuning offsets the Doppler shift of a counter propagating atom, $|\Delta_{pc}|-k_xv\approx0$.   When we account for the scattering from the counter propagating transverse beam which is of detuning $|\Delta_{pc}|+k_xv$, we find the optimal pump-cavity detuning to be $|\Delta_{pc}|\approx\kappa+k_xv$.  In other words, the pump beams should be chirped towards smaller detunings as the velocity distribution is compressed.  

Like free-space Doppler cooling, the velocity capture range of cavity cooling is limited.  From Eq{.}~\ref{dopplercooling} we can see that the capture range is limited to $\sim$$\kappa$ about a center frequency, $kv_0$.  The maximum cooling force on an atom moving in the 2D plane is~\cite{Vuletic01}:  $\mathbf{f}_{\mbox{max}}\cdot \mathbf{v}_0=-\hbar\Gamma_c(v_0)|(\mathbf{k}_x\mp\mathbf{k}_z)\cdot \mathbf{v_0}|$.  By balancing the cooling rate with recoil heating, one obtains a temperature limit proportional to $T_f\propto\kappa(1+C^{-1})$.  Note that for narrow cavities and light molecules such as OH, the recoil limit, $T_{rec}=\hbar\omega_{rec}/k_B$, can be larger than the cavity cooling limit, $T_f$.  The final temperature will therefore be limited by recoil.  For two-level atoms, we see that one does not need to be in the strong coupling regime to achieve cooling.  Cooperativity less than unity only slows the cooling rate and prevents the system from achieving the lowest possible temperature, set by $\hbar\kappa$.  This no longer holds true for the case of molecules, and the strong-coupling limit is required to cool efficiently, as we will see in Section~\ref{molecule}.

\section{Cavity cooling in the strong coupling limit without saturation}
\begin{figure}[t]
\begin{center}
\scalebox{0.6}[0.6]{\includegraphics{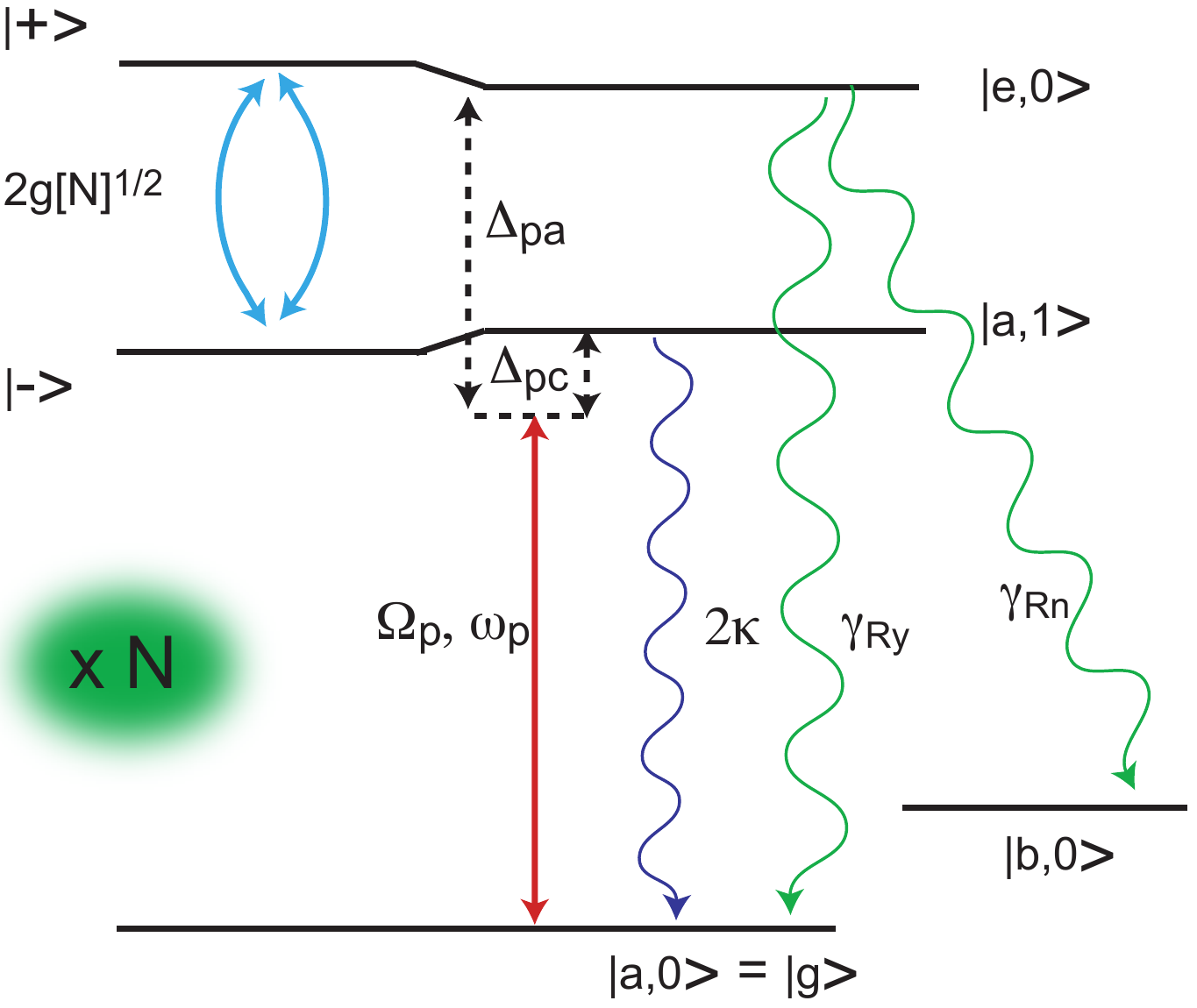}}
\caption{\label{fig:dressedstates}Dressed-state spectroscopy for N three-level atoms (molecules) coupled to a cavity mode in the presence of a transverse pump field, $\Omega_p$.  The detunings are set to $\left[\Delta_{pa},\Delta_{pc}\right]<0$ for cavity cooling, and the pump field connects the absolute ground state of the joint system $|g\rangle$ with $|e,0\rangle$.  For efficient cavity cooling, the magnitude of $\Omega_p$ is set so that the population of $|e,0\rangle$ is low, suppressing both incoherent scattering to $|g\rangle$ and Raman scattering to the secondary ground states, collectively represented by $|b,0\rangle$.  With these detunings, the atom(s)-cavity coupling, $g$, mixes the excited states to produce the dress-states $|\pm\rangle$.  The lower dressed-state $|-\rangle$, may be viewed as a polariton excitation, as it is mostly comprised of the cavity excitation.}
\end{center}
\end{figure}

In the good cavity limit ([$\gamma_{\perp},g]>\kappa$) or, more restrictively, the strong coupling regime ($g>\left[\kappa,\gamma_{\perp}\right]$), the perturbative treatment described above becomes less applicable since an intracavity photon can be reabsorbed coherently by the atom many times before finally being dissipated via the cavity mirrors or spontaneous emission.  The atom and cavity system can no longer be treated independently and it is advantageous to study the system in the dressed-state picture as shown in Fig{.}~\ref{fig:dressedstates}.  In this figure, the state space of the first excited states of the joint system, $|\pm\rangle$, is being probed by the pump laser of Rabi frequency $\Omega_p$.  The addition of the secondary ground state $|b,0\rangle$ accounts for all of the Raman loss channels found in realistic molecules, and is discussed more fully in Section~\ref{molecule}.  The inclusion the factor of $\sqrt{N}$ for the $N$ copies of the intracavity particles will be discussed in the context of multiparticle effects in Section~\ref{multiatom}.  The pump laser can only excite the $|e,0\rangle$ level since the transition to $|a,1\rangle$ is not electric dipole allowed.  At low saturations, $s\leq1$, this state decays via coherent elastic Rayleigh scattering, $\gamma_{Ry}$, to the ground state, which leads to heating.  Additionally, inelastic Raman scattering, $\gamma_{Rn}$, may occur, which depopulates $|g\rangle$.  The coherent atom-cavity coupling exchanges population from the atomic excited state $|e,0\rangle$, to the cavity's excited state $|a,1\rangle$, at the ``Rabi" frequency $2g$.  Thus, population can be decoupled from atomic decay if $g>\gamma_{\perp}$, and $|a,1\rangle$ will decay via mirror leakage at the rate $2\kappa$, which leads to cooling. 

An alternative, but wholly equivalent method for deriving Eqs{.}~\ref{cavscat1} and~\ref{atomscat} is to view the pump field as spectroscopically probing the dresses states $|\pm\rangle$, which are found by diagonalizing the Hamiltonian in Eq{.}~\ref{H_1atom1mode}.  Under the conditions of Eq.~\ref{conditions}, the states are: \bea |+\rangle&=&c_e|e,0\rangle+c_a|a,1\rangle \\  |-\rangle&=&c_a|e,0\rangle-c_e|a,1\rangle,\eea with \bea |c_e|^2&=&1-g^2/\Delta^2_{pa}, \\ |c_a|^2&=&g^2/\Delta^2_{pa}.\eea  The decay and populations of $|\pm\rangle$ may easily be obtained---with the justifiable assumption that the coherence between them vanishes rapidly---by treating them as independently pumped by $\Omega_p$.

Energetically, the cooling arises in a manner similar to that described in Section~\ref{perturb}, but whereas the sign of $\Delta_{pa}$ was not crucial to cooling in the weak coupling regime~\cite{Vuletic00,Vuletic01}, in the good cavity limit, cooling without atomic saturation is best when $\Delta_{pa}<0$ as shown in Figs{.}~\ref{fig:energy} and~\ref{fig:dressedstates}.  Furthermore, the damping of motion along the cavity axis in this regime is instead due to a cavity Sisyphus-like effect which arises from the time-delayed cavity field with respect to the atomic motion~\cite{Ritsch02_trans}.  Essentially, the dressed-state potential $|-\rangle$ is steeper as the atom climbs than when falling down the potential, which provides damping along the cavity axis.  This effect is also found when cooling via driving the cavity mode itself~\cite{Ritsch97,Ritsch98,vanEnk01}, though the role of the detunings is reversed and the scaling of the cooling rate with $N$ is less favorable~\cite{Ritsch01_drive}.  Additionally, a Doppler-like cooling mechanism along the pump field axis has been identified---referred to as ``polariton" cooling because the cooling arises from driving the $|-\rangle$ state---in which a strongly-coupled atom is cooled when its motion is perpendicular to the cavity axis~\cite{Ritsch05_polariton}.  Together with the cavity Sisyphus cooling, damping of the particle's motion in 3D is again possible as in the perturbative case.  As noted in Ref{.}~\cite{Murr06PRL}, the capture range can be enhanced in this regime by a factor of nearly 10 if one is willing to have the temperature limited by $\hbar\gamma/k_B$ rather than $\hbar\kappa/k_B$.  This is a small price as the temperature of ground state, polar molecular samples is currently limited to $T\geq100\hbar\gamma/k_B$.

Refs{.}~\cite{Ritsch02_trans} and~\cite{Ritsch03_review} review these effects and  connect how the cooling mechanism evolves from weak to the strong coupling regimes. 
Tractable diffusion and damping expressions are obtained in these works by keeping only lowest-order terms in $kv/\kappa$. More recent work extends this to all orders in velocity, completing the analysis of all the single-atom cavity cooling regimes~\cite{Murr06PRA_a,Murr06PRL,Murr06PRA_b}.

\begin{figure}[t]
\begin{center}
\scalebox{0.47}[0.47]{\includegraphics{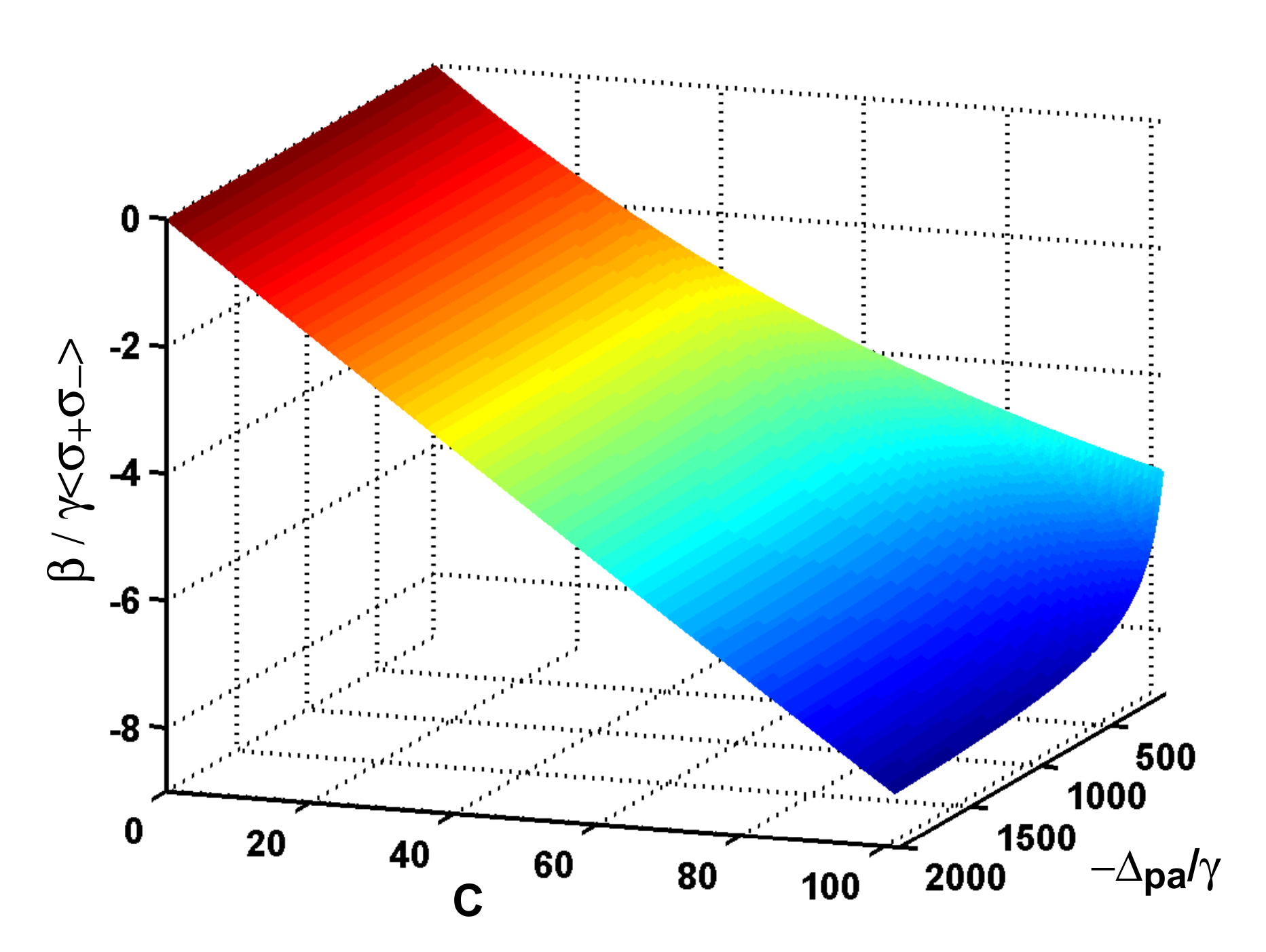}}
\caption{\label{fig:andras} Ratio of the cooling rate to the spontaneous emission rate, $\beta/(\gamma\langle\hat{\sigma}_+\hat{\sigma}_-\rangle)$, in which a negative-valued ratio indicates cooling. Parameters chosen for an OH molecule electronically excited in the cavity described near the end of Section~\ref{multimode} which has a finesse of 5000 and length of 2 cm.  The detuning $\Delta_{pc}$ is set such that the pump field is $-\kappa$ detuned from the lower dressed state.}
\end{center}
\end{figure}

A more sophisticated treatment of the cooling rate confirms the role of cooperativity as a benchmark to whether the particle can cool before spontaneously scattering.  Specifically, the ratio of the average velocity damping rate, $\beta$, to the free-space scattering rate is~\cite{Andras04}: \be \frac{\beta}{2\gamma_{\perp}\langle\hat{\sigma}_+\hat{\sigma}_-\rangle}=\frac{\omega_{rec}}{\gamma_{\perp}}\frac{\mathit{Im}\left\{(D^*)^2(z^2_c-g^2)\right\}}{|D|^2|z_c|^2},\ee where $z_c\equiv-\kappa+i\Delta_{pc}$,  $z_a\equiv-\gamma+i\Delta_{pa}$, and $D\equiv z_c z_a + g^2$.  Figure~\ref{fig:andras} plots this ratio versus the cooperativity and the detuning from atomic resonance.  The equations for the ratio of cooling to spontaneous emission in the cavity pumping case (i.e., $\Omega_d\neq0$ and $\Omega_p=0$) are cumbersome and not listed here.  In both cases, the ratio is less than unity when $C\leq1$, as expected.  This holds true for cooling to the blue of the cavity resonance as well (see Ref.~\cite{Ritsch03_review} for information on blue-detuned cooling).

\section{Three-level atoms representing molecules}\label{molecule}

As discussed in the last section, single two-level atom cavity cooling is well understood.  However, the efficacy of these cooling mechanisms comes into question when the scatterer has a more complicated internal structure.  It is the central point of this Article to extend the previous sections' analyses to the case of cooling molecules, and show that $C>1$ is an additional, necessary requirement for efficient cooling.

We may investigate multi-level scatters, such as molecules, by bundling all higher states into a single state denoted $|b\rangle$, and assume that the pump laser is so far detuned from $\omega_b$ that this level can never be depopulated once a single Raman scattering event has occured.  In other words, no coherence is developed between $|b,0\rangle$ and either $|a,0\rangle$ or the dressed-states $|\pm\rangle$ (see Fig{.}~\ref{fig:dressedstates}).  With this approximation, we need only to add an extra decay term to the master equation, Eq{.}~\ref{2levelmaster}, to incorporate the third level and, consequently, the set of non-ground states from which population effectively never returns:   \bea\label{3levelmaster}
\dot{\rho}&=&-i\left[\hat{H},\rho\right]+\gamma_{Ry}(2\hat{\sigma}_-\rho\hat{\sigma}_+-\hat{\sigma}_+\hat{\sigma}_-\rho-\rho\hat{\sigma}_+\hat{\sigma}_-)/2 \nonumber \\
& &-\gamma_{Rn}\left[\hat{\sigma}_+\hat{\sigma}_-,\rho\right]/2+\kappa(2\hat{a}\rho\hat{a}^\dagger-\hat{a}^\dagger\hat{a}\rho-\rho\hat{a}^\dagger\hat{a}).\nonumber \\
& &
\eea

For atoms such as cesium, only one additional laser is required to repump the atom back to its ground state where the cooling laser, associated with the pump field $\Omega_p$, can continue the cooling process.  Molecules, on the other hand, need more than one repumping laser to prevent population loss among the myriad vibrational, rotational, and spin-orbit levels.  Unfortunately, the generally low free-space branching ratio, $\Upsilon=\gamma_{Ry}/\gamma_{Rn}$, of Rayleigh-to-Raman scattering for molecules results in population shelving after only a few photon scattering events, thereby prematurely quenching the cooling process.  This ratio is fixed by the molecular structure, and it cannot be modified by detuning except when using detunings that are incredibly large $\Delta_{pa}\approx\omega_{b}$~\cite{CohenTannoudji}.  In other words, even though the molecule is unsaturated and coherently scatters the pump field, there remains a fixed---and generally large---probability for Raman loss per Rayleigh scattering event, regardless of how large the detuning.  Appendix~\ref{OH_transitions} lists the candidate cooling transitions for the radical OH, along with the respective $\Upsilon$'s.   

While building several repumpers becomes prohibitively complex and expensive, one may use a very far-off resonance transverse pump laser to simultaneously address all higher lying metastable states, as was described recently in Ref{.}~\cite{Pinske07}.   However, doing so naturally decreases the cooling rate by the ratio $\Omega^2_p/\Delta^2_{pa}$ as in Eq{.}~\ref{cavscat1}.  This low cooling rate in the far detuned case makes practical implementation difficult because of the short interaction time between hot molecules and the small cavity mode volume.   One can attempt to compensate by using a multimode cavity, as explored in Section~\ref{multimode} and Refs{.}~\cite{Vuletic01,Pinske07}.  The only solution for preventing population shelving is to ensure that there is a vanishingly small probability that the molecule will Raman scatter during the cooling time.  Because $\Upsilon$ is typically no larger than unity for molecules, there is a roughly 50\% chance that the molecule will be shelved in a dark state after each spontaneous scattering event.  It is therefore imperative that the excited state remain sufficiently unsaturated during the cooling process so that no spontaneous free-space scattering is likely to occur.  Cavities useful for cooling large samples occupy too small a solid angle for appreciable Raman suppression.  We will discuss these practical issues in more detail in Section~\ref{OHpracticalities}. 

While $\Upsilon$ cannot be modified (even by large detuning), the rate of scattering into the cavity, and thus the cooling rate, can be made to be larger than the Raman scattering rate by a factor equal to $(1+\Upsilon)C$.  The ratio $\Upsilon$ is for most molecules no larger than approximately unity.  Therefore, the rate of scattering into the cavity versus Raman loss is approximately equal to the cooperativity, $C$.  In other words, with respect to the lower dressed-state in Fig{.}~\ref{fig:dressedstates}, the cavity decay channel ($\propto 2\kappa$) can be made dominant over the atomic decay channels ($\propto \{\gamma_{Ry},\gamma_{Rn}\}$) if only one can achieve $C>1$.  Raman scattering is not simultaneously enhanced by the cavity because its frequency is far-detuned from the cavity resonance and does not spectrally fit into the cavity bandwidth.  This situation is markedly different from that of cavity cooling two-level atoms, where efficient cooling could be achieved even if the cooperativity is less than unity.  The cooling rate is slower, but the atoms are not shelved into dark states as occurs with molecules.  This is the essential difference between cavity cooling atoms and molecules, namely, molecules require $C\gg1$.  In the following sections, we will discuss methods to enhance the cooperativity for practical implementations of molecular cavity-assisted laser cooling.

\section{Enhancing single molecule cooperativities with multimode cavities}\label{multimode}

One method that can be used to achieve larger cooperativities is to decrease the volume of the cavity while simultaneously increasing its finesse.  This is the route taken by single atom cavity QED research~\cite{BermanBook94,Mabuchi02}, but is not suitable here as one wants to cool large diffuse samples several millimeters in width and would not fit into the required sub-millimeter sized cavities.  Moreover, to maintain a high $g^2/\kappa$ ratio, one needs to increase the finesse to the $10^4$ or $10^5$ regime.  Unfortunately, the strongest molecular electric dipole transitions (in light molecules like OH) are typically in the blue to UV wavelengths, and cavity finesses much greater than $10^3$ are currently commercially unavailable in the near UV. 

The geometry of the cavity is primarily constrained by the molecular beam size or trapped cloud dimension and by the inability to obtain high-finesse mirrors in the UV.  
Moreover, there is a trade-off between cooling volume and $g$:  a large cavity waist provides a large cooling volume, but $g$ will be correspondingly smaller.  In the case of OH produced by a Stark decelerator, the cavity length must be at least $\geq5$ mm to ensure that most of the molecules are enveloped by the mirrors, but $\leq20$ cm to fit inside typical vacuum chambers.  For OH excited on the P$_1(1)$ transition (see Appendix~\ref{OH_transitions}), $C=9.4\times10^{-2}$ for the TEM$_{00}$ mode of a cavity of radius of curvature (R) and length (L) approximately equal to 2 cm and finesse $F=5000$.  This cooperativity is certainly not sufficient for ensuring that the cooling rate will dominate the Raman loss rate.  

One may wonder whether cooling on purely vibrational transitions is more feasible given the high finesse coatings available in the infrared and the relatively few number of repumping lasers required to close the transition.  Unfortunately, the low decay rate of the vibrational transitions mitigates their utility for realistic cooling in all but long lifetime traps (see Section~\ref{OHpracticalities}).  For instance, OH's first vibrational transition at $\lambda=2.8$ $\mu$m, which naturally has the largest $\Upsilon$---equal to 1.6---has a slow, $\gamma=2\pi\cdot 2.7$ Hz, decay rate~\cite{Meijer05}.  With a single mode $R\approx L=1$ cm, $F=10^5$ cavity, the cooperativity is $C=34$, and the best achievable rate of scattering into the cavity would be $C\cdot\gamma=2\pi\cdot 100$ Hz.  The first vibrational overtone, at 1.4 $\mu$m, possesses a lifetime roughly a factor of two smaller, but has a wavelength at which it is much easier to obtain high power lasers.  A similar cavity would give $C=17$ and a maximum scattering rate into the cavity of $2\pi\cdot 23$ Hz.  However, in both cases the cavity waist is only $\sim$60 $\mu$m, which decreases the time fast molecules would spend in the cavity mode.

Before examining how multimode cavities can help increase $C$ for electronic transitions, we note that increasing the single-particle cooperativity by ``seeding," i.e., driving the cavity with a nonzero $\Omega_d$, does not help to stimulate more photon exchange from the pump beam to the cavity mode.  The same number of photons would be stimulated back from the cavity to the pump beam, thereby canceling the energy loss and adding recoil heating and atomic saturation.

Given these restrictions on minimum cavity length and maximum finesse, another method for increasing single molecule cooperativity is to increase the number of cavity modes available.   One could do this by wrapping more cavities around the molecule, but this is highly impractical.  Near-degenerate cavities---such as confocal, near-planar, and concentric---offer an ideal solution~\cite{Vuletic01} in that they can support many modes within the cavity linewidth $\kappa$.  The molecule can now scatter blue-shifted photons into many modes, effectively increasing $g$ by a large multiplicative factor, $\neff$.  

Before estimating this factor, we first incorporate a multimode cavity into the master equation:
\bea\label{3levelConfocal}
\dot{\rho}&=&-i\left[\hat{H},\rho\right]+\gamma_{Ry}(2\hat{\sigma}_-\rho\hat{\sigma}_+-\hat{\sigma}_+\hat{\sigma}_-\rho-\rho\hat{\sigma}_+\hat{\sigma}_-)/2 \nonumber \\
& &-\ \ \gamma_{Rn}\left[\hat{\sigma}_+\hat{\sigma}_-,\rho\right]/2 \nonumber\\
 & &+\ \ \sum_{i=1}^{M} \kappa_i(2\hat{a_i}\rho\hat{a_i}^\dagger-\hat{a_i}^\dagger\hat{a_i}\rho-\rho\hat{a_i}^\dagger\hat{a_i}), \\
\hat{H}&=&-\Delta_{pa}\hat{\sigma}_+\hat{\sigma}_--\Delta_{pc}\sum_{i=1}^{M} \hat{a}_i^\dagger\hat{a}_i+\Omega_p[\hat{\sigma}_-+\hat{\sigma}_+]/2\nonumber\\
& &+\ \ \sum_{i=1}^{M} \left\{ g_i\left[\hat{a}_i^\dagger\hat{\sigma}_-+\hat{\sigma}_+\hat{a}_i\right]+ \Omega_d[\hat{a}_i+\hat{a}_i^\dagger]/2\right\}.
\eea
We assume that all $M$ cavity modes within the bandwidth $\kappa$ couple to the atom so that we may remove the $i$ indices---effectively treating the modes as a single super-mode---and take $\sum_{i=1}^{M}g_i= \neff\ g$.  However, the coupling may not be equal for each mode at each position of the scatterer.  We next examine how the magnitude of $\neff$ may be calculated.

{Vuleti\'{c}} {\it{et al.}}~\cite{Vuletic01} and others previously~\cite{Heinzen87a} have shown that confocal cavities provide a better compromise between cooling rate and cooling volume over the other cavities (though the concentric has superior cooling rate performance~\cite{Vuletic01}).  Consequently, we will focus solely on the confocal cavity geometry. For this geometry, $\kappa=\pi c/2RF$, where the length and radius of curvature are equal, $L=R$, and the cavity waist is simply $w_0=\sqrt{R/k}$~\cite{Siegman}.   Realistic cavity mirrors have spherical aberration which limits the gains otherwise achievable with a confocal cavity.  Nevertheless, as long as $F<kR$, using a confocal cavity does increase the cooperativity over the single mode case.  In the UV, with the largest finesse one could hope to obtain, $F\approx5000$, the cavity length must be less than 1 mm for a single-mode cavity to be optimal, which is too small for accommodating the diffuse molecular cloud.  

The cooperativity of a single-mode cavity, $C=2F\Delta\Omega/\pi=6F/\pi kL$, can be enhanced to the following value, which is limited by spherical aberration ($sa$) (assuming a dipole scatterer oriented perpendicular to the cavity axis): \be C_{sa}=3F\Delta\Omega_{sa}/4\pi^2=3\sqrt{2F/\pi k R}.\ee  The following ratio provides an estimate of the effective number of additional modes:  \be\label{neff1} \neff=\sqrt{\frac{C_{sa}}{C}}=\left(\frac{\pi kR}{2F}\right)^{1/4}.\ee  In the above, $R$ is equal to the length of the confocal cavity and $F$ is the cavity finesse.  The solid angle of the confocal cavity---which is still much less than unity for cavities of interest---may be related to that of a single mode cavity of equal length and $F$ by: \be \Delta\Omega_{sa}=8\neff\Delta\Omega/3.\ee  For $F=5000$, $R=10$ cm, and $\lambda=308$ nm---parameters suitable for first electronic transition in OH---the effective enhancement is $\neff=5$.  As a concise figure of merit, one would like to have a confocal cavity that simultaneously maximizes spherical aberration-limited cooperativity and cooling area near the cavity waist, $A\approx\pi w_0^2$.  The product is \be C_{sa}A=\sqrt{\frac{18FR}{\pi k^3}}.\ee  We see from this expression that the optimal mirror quality and geometry is sensitive only to the product of $F$ and $R$, and a longer confocal cavity is favorable for fixed finesse.  This expression is not entirely fair, however, because a confocal cavity's mode volume is not limited by the $TEM_{00}$ waist, $w_0$, but rather the convolved waists of all the accessible modes.  For a spherically aberrated confocal cavity, this waist is $w_{sa}=2R_{sa}=2(2\pi R^3/kF)^{\frac{1}{4}}$~\cite{Vuletic01}, whose ratio to $w_0$ is: \be\label{conwaist} \frac{w_{sa}}{w_0}=2(kR/F)^{\frac{1}{4}}. \ee  This is always greater than unity for useful confocal cavities.  

In addition, these figures of merit do not factor in cooling time, proportional to $\kappa$ (see Eq{.}~\ref{cavscat1}), which is important for a non-stationary molecular sample.  We will discuss this consideration more in Section~\ref{OHpracticalities}.
Wavelength is not generally a tunable parameter, but redder transitions are favorable.  For the P$_1(1)$ OH transition listed in Appendix~\ref{OH_transitions}, a $L=R=2$ cm, $F=5000$ cavity gives $[C_{sa},\kappa,g_0, w_0, w_{sa}]=[1.1, 2\pi\cdot7.5\times10^5\mbox{ Hz}, 2\pi\cdot9.0\times10^4\mbox{ Hz}, 30\ \mu\mbox{m}, 0.3\mbox{ mm}]$.  Using the confocal cavity, a factor of $\sim$12 has been gained versus the single mode case (for which $C=0.09$), but even with the aid of a confocal cavity, one still cannot achieve a cooperativity much greater than unity for cavities accommodating samples of OH.  For a $R=L=10$ cm cavity, which would be better for molecular sample insertion and which also may be easier to obtain at a finesse as high as $F=5000$ in the UV, the cavity parameters become $[C_{sa},\kappa,g_0, w_0, w_{sa}]=[0.47, 2\pi\cdot1.5\times10^5\mbox{ Hz}, 2\pi\cdot1.8\times10^4\mbox{ Hz}, 70\ \mu\mbox{m}, 1\mbox{ mm}]$.

In practice, this $C_{sa}$ is most likely an upper bound.  In the experiments of {Vuleti\'{c}} {\it{et al.}}~\cite{Vuletic03_near, Vuletic03_far, Vuletic_Chan03}, the cavity happened to be misaligned from perfect confocality by $\sim$20 $\mu$m in the plane perpendicular to the cavity axis.  This splits the degeneracy of the cavity modes over a bandwidth of 200 MHz from the $TEM_{00}$ mode position.  Consequently, for their $R=7.5$ cm, $F=2000$ near-confocal cavity at 852 nm, $\neff$ should have equaled $\sim$4.5, but was experimentally found to be between 2.7 to 3.2, or 60-70\% of the expected value at the frequency of maximum mode density~\cite{Vuletic_Chan03}.  If we take this reduction as a pessimistic bound, then one can expect a cooperativity of $C_{sa}\approx[0.73,0.31]$ for the OH cavity cooling system of cavity length 2 cm and 10 cm, respectively.  The mode volume was 200 times larger than that of the $TEM_{00}$ mode~\cite{Vuletic_Chan03}, which is 95\% of what Eq{.}~\ref{conwaist} would predict for an ideal, spherically-aberrated confocal cavity.

The multimode enhancement, $\neff$, may be more accurately calculated by taking into account the actual confocal mode structure, as noted in Ref{.}~\cite{Ritsch02_trans}. This is done analytically using modes that are effectively uniform over the cavity length and by employing a numerical calculation that includes the Gouy-phase term.  This results in the following expression for $\neff$: \be\label{neff2} \neff=\frac{(2M'+1)!!}{(2M')!!},\ee where $2M'$ is the maximum mode index and $M=(M'+1)^2$ are the total number of modes seen by the scatterer.  Until the paraxial approximation breaks down for very large $M$, i.e., when $M\lambda\gg L$, $\neff$ grows roughly linearly with mode number as we expect from the previous discussion.  The number of modes supported by the cavity of {Vuleti\'{c}} {\it{et al.}}~\cite{Vuletic_Chan03} was measured to be 220, which produces $\neff=4.3$ from using Eq{.}~\ref{neff2}.  This is consistent with Eq{.}~\ref{neff1}'s value of 4.5, and the experimental near-confocal cavity realizes $\sim$70\% of this enhancement.  

For predictions of future cavity performance, Eq{.}~\ref{neff1} seems to be sufficient.  Using the numerical confocal cavity calculations, the authors of Ref{.}~\cite{Ritsch02_trans} reveal a constructive intracavity mode interference effect that reduces the temperature limit, $\hbar\kappa/k_B$, by as much as 20\% for scatterers offset from the cavity midpoint.  The cooling rate could be increased as well, though this effect would need to be confirmed with future simulations.  While multimode cavities can aid in increasing cooperativity, experimentally realistic confocal cavities cannot push $C$ much beyond unity.

\section{Complications and benefits of multi-intracavity scatterers}\label{multiatom}

As we have seen in the previous section, efficient molecular cavity cooling is not possible when $C<1$, as is the case for the P$_1(1)$ transition of OH when the cavity is made long enough to accommodate an experimentally realizable sample.  The Raman loss rate will dominate the cooling rate, prematurely quenching the cooling process.  For multiple intracavity molecules, cooling is still impossible as long as the molecules act independently of one another and $\Gamma_{c}$ is unmodified by many-particle effects.  In other words, because the Rayleigh-to-Raman scattering ratio, $\Upsilon$, will never be much larger than unity with no repumpers (or even with a single one), the $C>1$ regime is necessary to enhance coherent Rayleigh scattering into the cavity mode over the Raman free-space scattering.  To achieve this strong coupling in the absence of collective effects, one has no other recourse but to make small cavities, an impractical compromise if one intends to cool large samples.  
\begin{figure}[t]
\begin{center}
\scalebox{0.45}[0.45]{\includegraphics{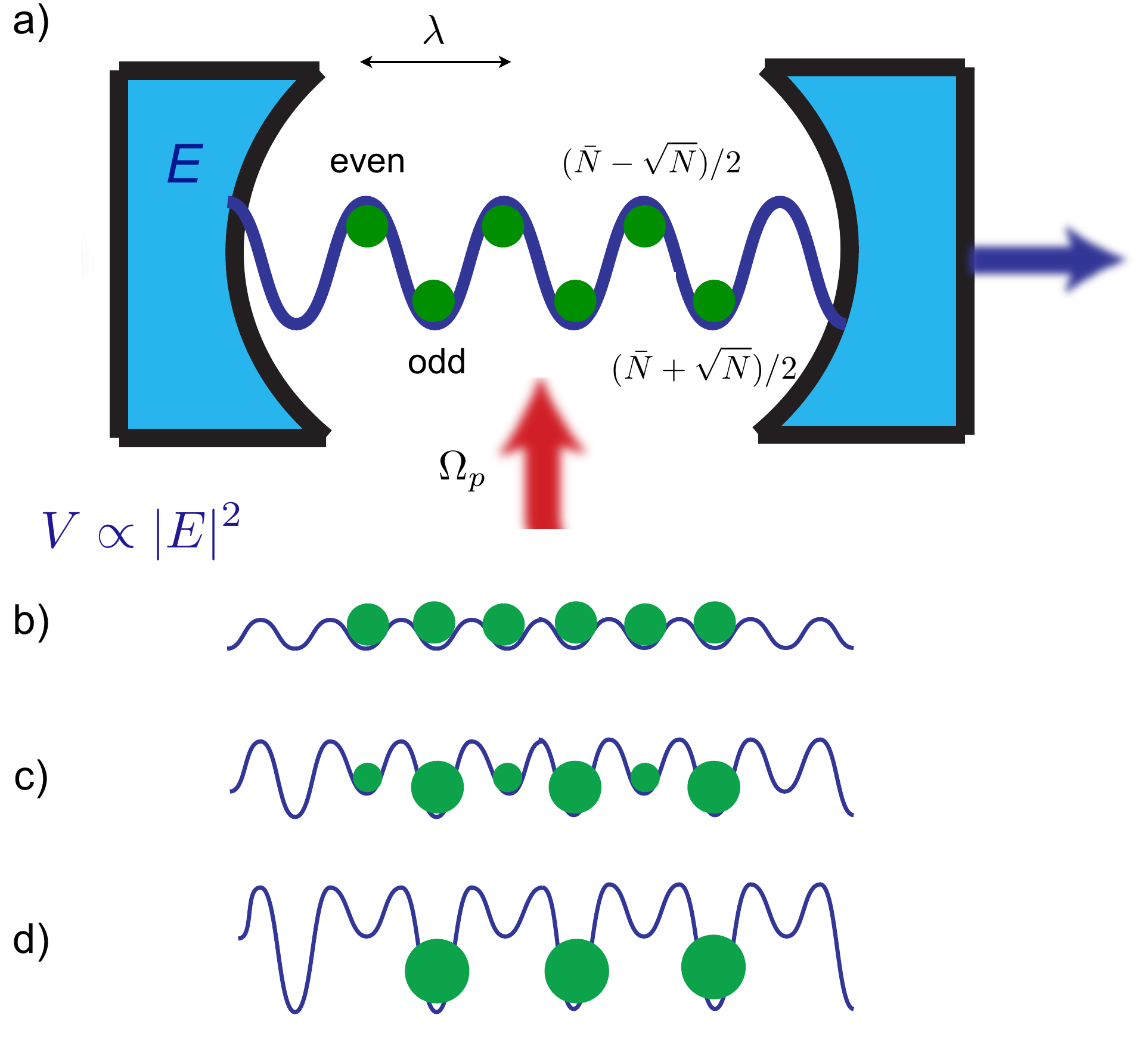}}
\caption{\label{fig:threshold} a) Cartoon of multiple intracavity particle scattering and self-localization.  Green dots represent the scatterers and the blue sinusoidal line is the intracavity field, $E$.  There are two antinodes per wavelength, distinguished by opposite, even or odd phases.  The even are represented as the upper antinodes while the odd are the lower.  b) Below a threshold transverse pump Rabi frequency, $\Omega_p<\Omega_{th}$, the atoms at the even and odd sites experience an identical dipole trapping force, proportional to $|E|^2$.  The random nature of the atom position implies that at any given time there could be $\sqrt{N}$ more atoms in the odd sites than the even.  c) This leads to an increase in scattered light from the unpaired ``odd" atoms, and the interference with the pump beam creates an ever deeper optical potential for atoms in the odd sites.  d) The even sites are depopulated over the odd, leading to a phase transition from particles located every antinode to a $\lambda$ spacing. Concomitantly, superradiance ensues, which increases the per atom cooperativity by $N$. The choice of odd over even is arbitrary and the symmetry is spontaneously broken in favor of one versus the other.}
\end{center}
\end{figure} 

The situation of multiple intracavity scatterers adds many-body complexity to the cooling physics, but also a means to increase the per-molecule cooperativity beyond unity.  To understand how this collective enhancement may arise, let us first consider $N$ randomly spaced scatterers (which we will presently refer to as atoms) inside a cavity, as depicted in Fig{.}~\ref{fig:threshold}.    For simplicity, we assume all $N$ atoms are located at the antinodes.  Atoms at the nodes do not couple to the field ($\psi=0$) and those offset from both the node and antinodes have a diminished coupling ($\psi<1$).  We can renormalize $N$ to $\Neff=\sum_i^N\psi_i$ and consider the $\Neff$ atoms equally distributed among the antinodes (we will drop the {\it{eff}} subscript from this point forward).  There are two antinodes per wavelength and the electric field oscillates exactly $\pi$ out of phase between them.  We designate every other antinode as ``even" or ``odd" to highlight this phase difference.  Figure~\ref{fig:threshold} depicts the even and odd antinodes, along with the on average $\bar{N}/2$ atoms in the even sites and $\bar{N}/2$ in the odd.  In the absence of motional fluctuations, there are exactly the same number of atoms at the two types of antinodes.   For every ``even" atom that scatters a photon in the cavity mode from the pump beam, there will be another ``odd" atom that scatters a $\lambda/2$-displaced photon along the cavity axis.  These two photons destructively interfere, preventing cavity field build-up.  

The atoms have a finite temperature, however, and statistical fluctuations will cause momentary imbalances in the particle number at the even versus odd sites.  Assuming Poissonian fluctuations about the mean, at any given moment there are $(\bar{N}+\sqrt{N})/2$ atoms in the ``odd" sites versus $(\bar{N}-\sqrt{N})/2$ in the even.  Of course, the choice of excess atoms in the odd sites is arbitrary, but we will assume this for concreteness.  These additional $\sqrt{N}$ atoms in the odd sites are unpaired by any atoms located at integer multiples of $\lambda/2$ and therefore scatter photons into the cavity without destructive interference.  Thus, a cavity field is built-up with Rabi frequency proportional to $\sqrt{N}g$.  This produces a per atom cooperativity equal to: \be C=\frac{1}{N}\frac{(\sqrt{N}g)^2}{\kappa\gamma}=\frac{g^2}{\kappa\gamma},\ee which is equal to the single particle cooperativity.  

No multiparticle cooperative effect is seen in the cavity scattering rate---which is $N\Gamma_c$---even though the spectrum of the joint atoms-cavity system exhibits a eigenmode splitting proportional to $\sqrt{N}g$.  This highlights the difference between pumping the cavity mode itself ($\Omega_d$) versus pumping the atomic mode directly with a beam transverse to the cavity axis ($\Omega_p$).  In the latter case, the atoms act as independent scatterers when observing the rate at which photons are coupled to and then leaked from the cavity mode.  In the former, the atoms collectively act as a giant dipole that modifies the cavity transfer function (atoms-cavity spectrum).

The situation is dramatically modified if the pump is made stronger than a critical field $\Omega_p\geq\Omega_{th}$~\cite{Ritsch02_bragg, Ritsch05_threshold}.  A more detailed explanation will follow after this general overview.  Because the cavity field is red-detuned from the atomic resonance, an optical dipole trap may be formed with trap minima centered at the cavity antinodes.  However, an intracavity field will not be formed if there is no population imbalance between even and odd sites.  As the particle positions fluctuate, a population imbalance will form and a cavity field will be generated.  If this thermal position fluctuation is weaker than the ensuing optical dipole trap, then the imbalance between even and odd wells will grow in a runaway (positive feedback) process:  Eventually all the atoms will migrate to favored set of wells as this further increases the cavity field and the trap depth.  A lattice of periodicity $\lambda$ will form in every other antinode of the intracavity dipole trap whose strength is proportional to $|E|^2$ (see Fig{.}~\ref{fig:threshold}).  Since scattered pump photons into the cavity mode now {\it{constructively}} interfere, a large cavity field is built-up proportional to $Ng$.  The resulting collective single-particle cooperativity is now \be  C=\frac{1}{N}\frac{(Ng)^2}{\kappa\gamma}=\frac{Ng^2}{\kappa\gamma},\ee which is $N$ times the single-particle cooperativity, $C_N=NC$.  The total cavity emission is now superradiant, $N^2\Gamma_c$,~\cite{Dicke,Haroche83} and the per particle scattering rate into the cavity mode is collectively enhanced to $N\Gamma_c$.  Since one can place many molecules inside the cavity mode, the collective cooperativity can be much greater than unity, $C_N\gg1$.  Raman scattering can now be completely suppressed relative to the elastic scattering rate.  

\section{Threshold for superradiance}

The question remains how exactly to trigger this phase transition from a random particle distribution to one of self-organization. Note, this is a spontaneous symmetry breaking transition---initial fluctuations determine whether the particles localize at the even or odd sites.  We now revisit the formalism of Ref{.}~\cite{Ritsch05_threshold} to better understand the problem.  From a Hamiltonian that incorporates the weak driving approximation, Eq{.}~\ref{conditions}, we obtain the stochastic time evolution equation for the semiclassical cavity field: \bea \dot{\alpha}&=&i\left[\Delta_{pc}-U_0\sum_j\cos(kz_j)\right]\alpha \nonumber \\
&-&\left[\kappa+\Gamma_0\sum_j\cos^2(kz_j)\right]\alpha \nonumber \\ &-&\etaeff\sum_j\cos(kz_j)\cos(kx_j)+\xi_\alpha,\eea where $z_j$ and $x_j$ are the positions of the $j$th atom along the cavity field and pump field, respectively, and $\xi_\alpha$ is the Langevin noise term for the cavity field.  $U_0$ and $\Gamma_0$ are the dispersive drive and decay rates of the cavity field via the atomic medium, respectively: \be U_0=\frac{g^2\Delta_{pa}}{\Delta^2_{pa}+\gamma_{\perp}^2}, \ \ \ \Gamma_0=\frac{g^2\gamma_{\perp}}{\Delta^2_{pa}+\gamma_{\perp}^2}. \ee  The effective cavity pump strength due to coherent scattering of the pump field by the atom is: \be \etaeff=\frac{g\Omega_p}{-i\Delta_{pa}+\gamma_{\perp}}.\ee  Expressions $\etaeff^2/\kappa$ and $\Gamma_0$ reduce, respectively, to Eqs.~\ref{atomscat} and~\ref{cavscat1} when $|\Delta_{pc}|\approx\kappa$.

The force on the atoms along the cavity and pump axes, respectively, are:  \bea \label{forceeqn}
\dot{p}_{z_j}&=&-\hbar U_0|\alpha|^2\frac{\partial}{\partial z_j}\cos^2(kz_j) \nonumber \\ 
&-&i\hbar(\etaeff^*\alpha-\etaeff\alpha^*))\frac{\partial}{\partial z_j}\cos(kx_j)\cos(kz_j)+\xi_{z_j} \nonumber \\
\dot{p}_{x_j}&=&-\hbar U_0(\Omega_p/g)^2\frac{\partial}{\partial x_j}\cos^2(kx_j) \nonumber \\ 
&-&i\hbar(\etaeff^*\alpha-\etaeff\alpha^*))\frac{\partial}{\partial x_j}\cos(kx_j)\cos(kz_j)+\xi_{x_j}, \nonumber \\
&  &
 \eea where $\xi_{z_j}$ and $\xi_{x_j}$ are Langevin noise terms for the motion along the $\hat{z}$ and $\hat{x}$, respectively.  If we look at the sign of the interference term between the cavity and pump fields, proportional to $\etaeff$, a checkerboard-like pattern appears in the crossed intracavity and pump fields.  A single line of antinodes along the cavity axis is depicted in Fig{.}~\ref{fig:threshold}.  Moving one half-wavelength perpendicular to the cavity axis---along the pump field and in the $\hat{x}$ direction---changes the phase of the pump field while the cavity field phase stays constant:  The spacing of the even-odd antinodes are shifted by $\lambda/2$, thus creating the checkerboard pattern.  If the even and odd antinodes are equally populated, then the sum over $\cos(kz_j)\cos(kx_j)$ is zero and there is no contribution to the cavity field from the $\etaeff$ term.  Once the population becomes imbalanced, the cavity field is pumped by this term and grows in a runaway process as the atoms migrate to one set of antinodes.  
 
The optical trapping potential felt by each atom along a string of antinodes parallel to the cavity axis is \be V(z)=U_2\cos^2(kz)+U_1\cos(kz), \ee which may be understood from the first of Eqs{.}~\ref{forceeqn}, with $U_1$ the strength of the interference term between the transverse pump and cavity mode and $U_2$ the dipole trap from the cavity mode.  The depths of the potential are \bea U_2&=&\hbar I_0U_0N^2\left<\cos(kz)\right>^2 \\
 U_1&=&2\hbar I_0 N\left<\cos(kz)\right>[\Delta_{cp}-NU_0\left<\cos^2(kz)\right>],
 \eea where \be I_0=\frac{|\etaeff|^2}{[\kappa+N\Gamma_0\left<\cos^2(kz)\right>]^2+[\Delta_{cp}-NU_0\left<\cos^2(kz)\right>]^2}.\ee This latter expression may be understood as the per atom scattering rate into the cavity mode and is a generalization of Eq{.}~\ref{cavscat1} for $N>1$.  To understand why one set of antinodes are preferred over another, we look at the signs of $U_1$ and $U_2$.  When $U_0<0$, as is the case with negative detuning from the atomic line, the $U_2$ term represents a cavity field that functions as a red-detuned optical lattice with equal trap depth in the even and odd antinodes.  It is the $U_1$ interference term that breaks the symmetry between the antinodes and gives rise to the checkerboard pattern of atoms occupying only every other antinode.  We describe this process as follows:  The spatial average of atomic positions along the cavity axis, $\left<\cos(kz)\right>$, is either closer to +1 or to -1 initially.  If $\left<\cos(kz)\right><0$ initially, then there are more atoms near the odd sites and consequently, the odd sites have a deeper potential than the even.  More atoms localize in the stronger wells of the odd sites, further reinforcing the asymmetry between the odd and even antinodes.  This results in a runaway self-localization process, leading to $\left<\cos(kz)\right>$ assuming a value closer to -1.  The same holds true if $\left<\cos(kz)\right>$ is initially +1, except the atoms eventually localize around the even sites.   It has been shown that the final state---all atoms at the even checkerboard sites or all at the odd---is stable once formed~\cite{Ritsch02_bragg,Ritsch05_threshold}.  

We now address the criteria for reaching threshold, $\Omega^{th}_p$.  The threshold condition has been derived in two ways~\cite{Ritsch05_threshold}.  The first method employs a mean-field approximation that assumes an intracavity gas of constant density, which is valid in the thermodynamic limit $N\rightarrow\infty$, $g\rightarrow0$, $\kappa=\mbox{const}$, and $Ng^2\propto N/V=\mbox{const}$.  For driving the lower dressed state below resonance by an amount equal to $\kappa$, i.e., $\Delta_{pc}=NU_0-\kappa$, where $NU_0$ is the energy shift of the $|-\rangle$ dressed state, the threshold is:  \be\label{meanfieldthres} \Omega_p\geq\Omega^{th}_p=\sqrt{\frac{k_BT}{\hbar\kappa}}\frac{\kappa|\Delta_{pa}|}{\sqrt{N}g}\sqrt{2}.\ee  We see that the depth of the optical dipole trap at threshold scales linearly with temperature ($V_{th}\propto\Omega_{th}^2$), is scaled inversely by the temperature limit $T_f\approx\hbar\kappa/k_B$, and is inversely proportional to the rate at which pump photons are scattered into the cavity mode, $\propto Ng^2/\kappa\Delta^2_{pa}$. This may be simply understood from the statement that the trap depth must be large enough to quench the diffusion due to thermal energy.

A threshold that scales inversely with $\sqrt{N}$ would not necessarily prevent the triggering of self-localization in experimentally realizable samples of molecules, but numerical calculations~\cite{Ritsch05_threshold} indicate that a hysteresis effect in the phase transition pushes the onset out to larger pump fields.  The numerically verified threshold becomes: \be\label{numthres} \Omega_p\geq\Omega_{th}=\sqrt{\frac{k_BT}{\hbar\kappa}}\frac{\kappa|\Delta_{pa}|}{N^{1/4}g}\frac{\sqrt{\pi}}{2}.\ee  This significantly worse scaling with $N$ would prevent the triggering of the superradiance.

A major impediment to triggering threshold is the need to do so without saturating the transition.  In other words, one cannot simply increase $\Omega_p$ without a cost in the number of spontaneous Raman emissions.  While $\Omega_p$ must be greater than threshold:  \be\label{threscond} \Omega_p>\sqrt{\frac{k_BT}{\hbar\kappa}}\frac{\kappa|\Delta_{pa}|}{N^{1/x}g},\ee where $x$ might equal to 2 or 4, the saturation condition must additionally be satisfied:  \be  s\approx\frac{\Omega^2_p}{4\Delta^2_{pa}}\ll 1.\ee   This implies that the molecule number must be much greater than: \be\label{Nthreshold} N_0>\left[\sqrt{\frac{k_BT}{\hbar\kappa}}\frac{\kappa}{2g\sqrt{s}}\right]^x.\ee Using the following parameters for OH in the 2 cm confocal cavity described in Section~\ref{multimode} and assuming T=10 mK, we have $N_0>8.5\times10^3$ for $x=2$ and $N_0>7.3\times10^7$ for $x=4$.  As discussed in Section~\ref{OHpracticalities}, $N_0$ could be achievable for $x=2$ with improvements in Stark deceleration, but not likely achievable for $x=4$.  Thus, the question of whether the threshold scales as $N^{-1/2}$ or as $N^{-1/4}$ is of paramount importance.  

Seeding the cavity by driving the cavity mode itself with $\Omega_d\neq0$ is one possible method to increase the effectiveness of the transverse cavity cooling scheme, and the phase of the drive field with respect to the pump field has been previously shown to affect the symmetry breaking of the self-localization process~\cite{Andras04}.  Seeding introduces an intracavity optical dipole trap without having to first scatter from the intracavity atomic medium.  One might expect this to hasten the cooling process or decrease the number of molecules required for threshold, but neither effect has thus far been seen in quantum Monte Carlo simulations.  This may be due to the fact that a lattice formed by driving the cavity does not contribute to the positive feedback mechanism responsible for self-localization.  Nevertheless, any possible method for reducing the threshold requirements for localization warrants more investigation via additional numerical simulations. 

Reference~\cite{Ritsch04_suppression} identified a mechanism that suppresses free-space scattering well over what is expected from the $C_N$ factor, once self-localization has been triggered.  This is due to pump-cavity mode interference in which the collective atomic dipole oscillates out of phase with respect to the pump, suppressing the atoms' excitement.  This effect points to another method for achieving superradiance in the case where $N<N_0$: increase the saturation, $s$, as the molecules enter the cavity to lower the criteria for reaching threshold.  Once self-localization is established, spontaneous emission should be quenched, and the process only sacrifices a fraction of the initial molecules to achieve self-localization for the remaining ones.  For example, this technique may be employed for cooling on the vibrational transitions which would nominally require $N_0>10^7$ for $x=2$, but can only Raman scatter at a maximum rate $\gamma/2\approx2\pi\times50$ Hz.

The simulations in Ref{.}~\cite{Ritsch05_threshold} suggest that the threshold should scale with particle number as $x=4$, but the experimental results on self-localization~\cite{Vuletic03_far,Vuletic_Chan03,Vuletic04} are consistent with $x=2$.   References~\cite{Vuletic_Chan03,Vuletic04} attribute this experimental $x=2$ scaling to the onset of recoil induced resonances (RIR)~\cite{Berman99}, but it is noted in Ref{.}~\cite{Ritsch02_bragg} that RIR would lead only to transient enhancements as opposed to the long-lived collective states described by self-localization.  Future work will elucidate the dynamics of this phase transition with additional simulations.  Questions to be explored are: how the transition threshold scales with intracavity particle number; the role of seeding; the characteristic onset time for superradiance; and how many molecules are lost before spontaneous emission is suppressed in the self-localized state.  An additional scenario for simulations to explore is the case in which the recoil energy of the particle of interest, $\hbar\omega_{rec}$, is comparable to the final cooling temperature, $\hbar\kappa/k_B$.  All these questions are of crucial interest for understanding the efficacy of using transversly-pumped cavities to cool molecules.  This latter question is of particular interest for the relatively light OH molecule whose recoil energy, $12$ $\mu$K, is comparable to $\hbar\kappa/k_B=7$ $\mu$K for the 10 cm confocal cavity described above.   This naturally leads us to the last main section of this Article, experimental considerations for the cooling of molecules, and in particular, OH.

\section{Experimental considerations}\label{OHpracticalities}

\begin{figure}[t]
\begin{center}
\scalebox{0.535}[0.535]{\includegraphics{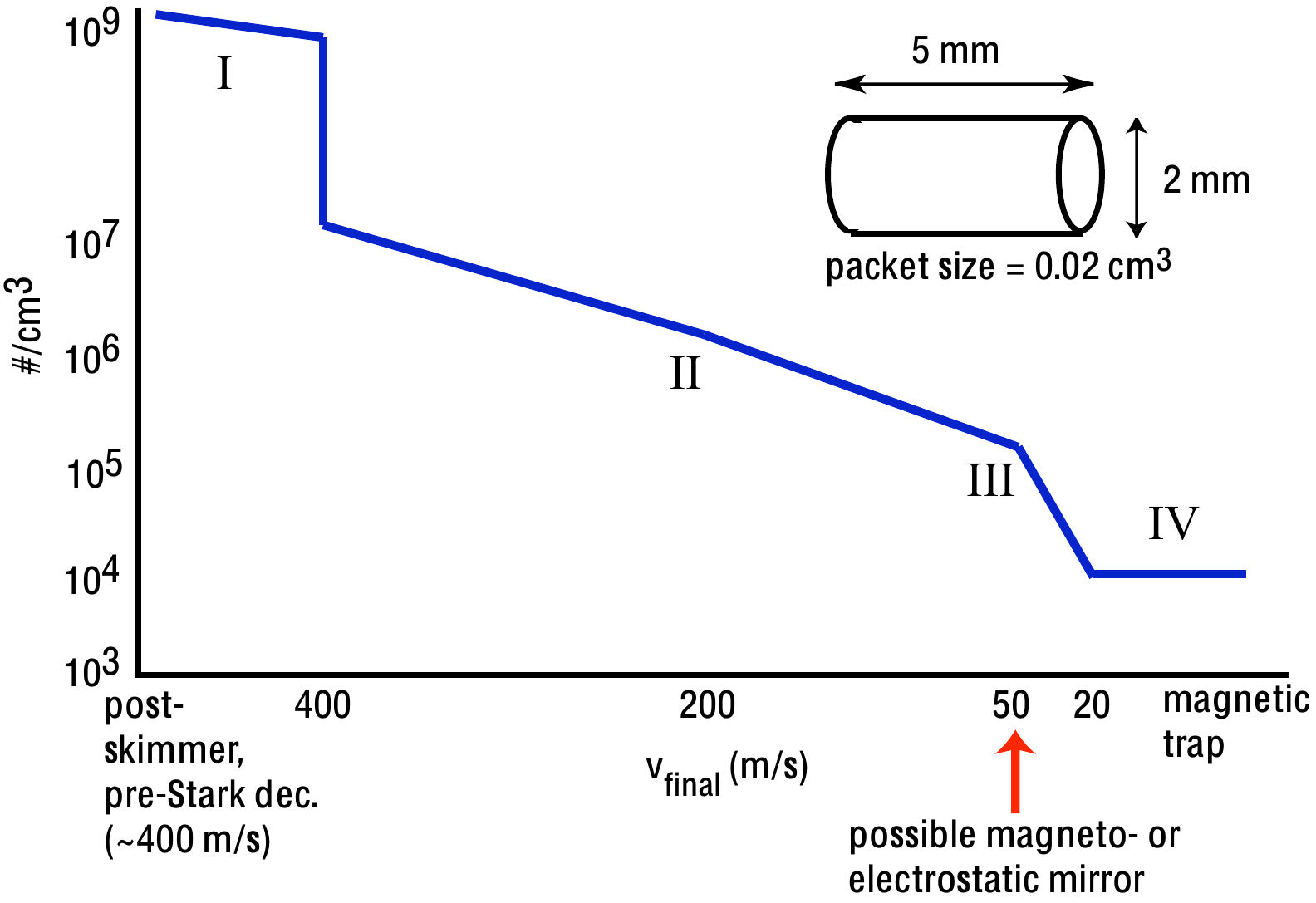}}
\caption{\label{fig:SDeff} Approximate experimental OH Stark-decelerator efficiency curve---density versus final velocity---for the JILA apparatus~\cite{Hudson04, Sawyer07b,Sawyer07}.  Inset shows the approximate volume of the OH packet in zones  {\textrm{II}} and  {\textrm{III}}. The packet is larger in the pre-Stark decelerator zone {\textrm{I}}, and expands to fill the magnetic or electric trap volume in zone {\textrm{IV}}.}
\end{center}
\end{figure} 
Most of the physics to be studied with polar molecules requires them to be in their rovibronic ground state.  Supersonic expansions of the molecules entrained in a buffer gas readily produces packets of polar molecules with quenched rotational and vibrational motion. The technique of Stark deceleration~\cite{Meijer99} can readily produce slow packets of these ground-state molecules, and in particular, the polar molecule OH~\cite{Bochinski03}.  OH is produced either via water discharge or photolysis of nitric acid. When entrained in Xe or Kr, nearly all of the OH---upon expansion---is in the lowest rovibrational $^2\Pi_{3/2}$ \LD ground states.  Although the packet that is formed has a much narrower velocity spread than expected from a Maxwell-Boltzman distribution, the mean center-of-mass (CM) velocity is approximately 400 m/s when using Xe as the buffer gas.  The packet has a $\sim$15\% and a $\sim$7\% velocity spread in the longitudinal and transverse dimensions, respectively.  Thus, in the moving frame, the molecular packet is cold (on the order of 1 K).  Before slowing the packet to near-zero CM velocity, the packet must pass through a skimmer to prevent backscattering from collapsing the supersonic expansion and to limit the flux of unwanted gas into the decelerator chamber.

A major practical difficulty with cavity-assisted laser cooling of molecules involves ensuring that a sufficiently large molecular sample experiences the mode volume of the cavity for a long enough period of time to cool.  At best, cavity waists are no larger than a millimeter, and ground state polar molecules have yet to be produced in sufficient number at CM velocities below $\sim$10 m/s.  For the technique of Stark deceleration, Figure~\ref{fig:SDeff} shows the inherent trade-off between slowed packet density and final velocity. 
\begin{figure}[t]
\begin{center}
\scalebox{0.53}[0.53]{\includegraphics{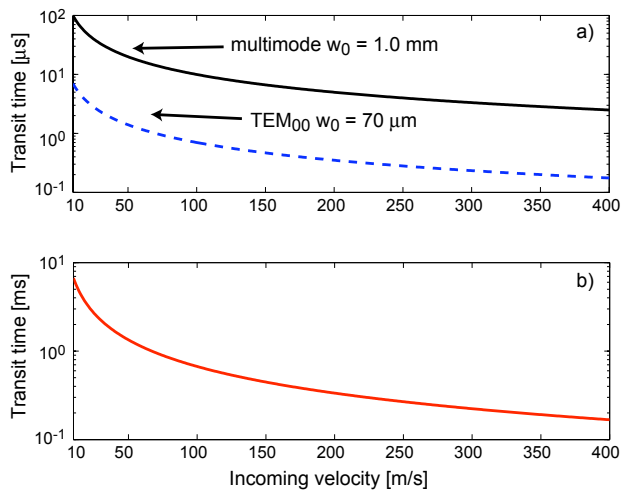}}
\caption{\label{fig:transittime} a) Transit time for crossing perpendicular to the cavity waist.  Black (solid) curve for crossing all the modes of a confocal cavity, blue (dashed) for crossing only the $TEM_{00}$ mode of the cavity.  b) Time to cross 66\% of the cavity length, $L=10$ cm, parallel to the cavity axis.}
\end{center}
\end{figure} 

The efficiency curve is divided into four zones, each zone being amenable to a different regime of cavity cooling.  Post-skimmer, the OH packet is typically of density 10$^8$ to 10$^9$ cm$^{-3}$, as shown in zone {\textrm{I}} in Fig.~\ref{fig:SDeff}.   Because the packet is moving at high velocity, it experiences the cavity mode for only a short time.  However, the number of OH in the cavity mode may be sufficiently large to trigger rapid cooling.  Self-localization happens no faster than $1/\kappa$, but seems to occur within tens of microseconds or less~\cite{Vuletic03_far,Ritsch05_threshold}.  Figure~\ref{fig:transittime} shows the time required to cross the cavity waist for a given velocity.  It may be possible to transversely cool the fast OH packet with a cavity.  Since Stark deceleration would be much more efficient with an OH packet of narrower transverse velocity spread~\cite{Sawyer07b}, the cooling in zone {\textrm{I}} could lead to the Stark slowing of many more molecules in zones {\textrm{II}}--{\textrm{IV}}.

Zones {\textrm{II}} and {\textrm{III}} offer a compromise between particle number and cavity transit time.  In zone {\textrm{II}}, it may be possible to increase the molecule-cavity interaction time by redirecting the molecule along the cavity axis with an electrostatic guide.  The molecules are slow enough in zone {\textrm{III}} to be stopped and reflected by an electrostatic~\cite{Meijer04mirror} or magnetostatic~\cite{Sawyer07} mirror.  During the reflection time, the molecules could spend more than a millisecond in the mode of an optimally situated cavity.  

Magnetic~\cite{Sawyer07} and electrostatic~\cite{MeijerETrap} traps and AC electric traps~\cite{Veldhoven05} have been used to trap polar molecules at the terminus of a Stark decelerator, in zone {\textrm{IV}}.  However, room temperature blackbody radiation limits the OH lifetime in these traps to $<3$ s~\cite{Meijerblackbody}.  Thus, the cooling must occur on a time scale much faster than this.  While 1 kHz cooling rates, for example, are fast compared to this lifetime, it must be noted that the fraction of the trap volume occupied by the cavity waist can be quite small.  Consequently, the molecules do not spend a large amount of time being cooled, and a much larger cooling rate than expected is necessary.

Future work--using numerical cavity cooling simulations incorporating molecular motion--will explore optimal cavity and decelerator geometries for cavity cooling OH.  The ground state polar molecular samples are currently 1000 times hotter than the cooling limit imposed by the cavity linewidth.  Simulations will address the efficiency and scaling of cooling such hot samples as well.  Future improvements to the Stark decelerator technique~\cite{Sawyer07b} will produce higher densities of slow molecules, seemingly an important step towards successfully achieving the cavity-assisted laser cooling of molecules. The use of other techniques, such as feedback~\cite{BarkerFeedback} and optical Stark deceleration~\cite{BarkerOStark} may be necessary for increasing the cooling rate and obtaining a larger number of slowed molecules. 

In summary, we have identified important necessary conditions for cavity cooling ground state molecules, whose open channels and high initial temperatures pose unique challenges.  We determine that to prevent Raman loss and thereby achieve efficient cooling, the cooperativity should be greater than unity.  Several methods for increasing the cooperativity and cooling rate are examined.  These include the use of multimode cavities; simultaneously driving (seeding) the cavity mode while transversely pumping the atomic medium; and inducing a self-localizatizing phase transition of the molecules' positions.  While multimode cavities are useful for increasing the cavity cooling volume and raising the cooperativity to near unity, only by inducing the superradiant phase transition can one achieve molecular cavity cooling with certainty.  We have assessed the feasibility of triggering this superradiant state as well as address the possibility of cooling under current experimental constraints, such as UV mirror coating technology, with particular emphasis given to the present performance of OH Stark decelerators.  More in-depth numerical simulations of the seeding and self-localization dynamics in the presence of molecule motion---either from a beam or due to a harmonic trap---are required to fully access the feasibility of efficient cavity-assisted laser cooing of molecules.  Future work will address these questions via both quantum Monte Carlo simulation~\cite{Ritsch03_review} and experiment.

\begin{acknowledgments}
We thank J. Dunn, C. Greene, J. Asb{\'{o}}th, and B. Stuhl for useful discussions, and acknowledge financial support from DOE, NIST, NRC, and NSF.  B$.$ Lev is a National Research Council postdoctoral fellow. 
\end{acknowledgments}
\begin{table*}[t]
 \caption{\label{OH}Comparison of the three best OH electronic cooling transitions.  P$_1$(1) has a better bare Rayleigh-to-Raman ratio, $\Upsilon$, but requires one microwave pumping stage and does not have a cycling hyperfine transition. The Q$_1$(1) transition has a smaller $\Upsilon$ and a cycling transition on $F''=1\rightarrow F'=2$ if two microwave pumping stages are used to prepare the OH in the  $F''=1$ hyperfine ground state.  A compromise transition is Q$_{21}$(1), which has an intermediate $\Upsilon$ but no cycling hyperfine transition.  Note that the P$_1$(1) transition is 92\% closed with just one repumper on the P$_{12}$(1) line. Transition wavelengths and lifetimes are from the software packages HITRAN~\cite{HITRAN} and Lifbase~\cite{Lifbase}.}  
\begin{ruledtabular}
\begin{tabular}{c|c|c|c|c|c|c|c|c} 
& $\lambda_{ae}$ (nm) & $[J', N']$ & $\Gamma/2\pi$ ($10^5$ Hz) & $\Upsilon$ & \# repumpers & $v''\neq0$ & $\Delta F=+1$ ? & \# $\mu$-wave pulses\\\hline P$_1$(1) & 308.256 & [1/2, 0] & 2.32 & 1.43 & 2 & 0.4\% & no & 1\\\hline Q$_1$(1) & 307.933 & [3/2, 1] & 2.32 & 0.28 & 4 & 4\% & yes & 2\\\hline Q$_{21}$(1) & 307.937 & [1/2, 1] & 2.32 & 0.65 & 2 & 1\% & no & 0
\end{tabular}
\end{ruledtabular}
\end{table*}

\appendix

\section{Atom-cavity coupling parameters}\label{cavityparameters}

This section describes the various parameters encountered when analyzing a cavity QED system and attempts to connect this language with that used by other authors. 
Single atom strong coupling requires the atom-cavity coupling, $g_0$, to be much larger than both the atomic dipole decay rate, $\gamma_\perp=\gamma/2$, and the decay rate of the cavity field, $\kappa$.  Specifically, the saturation photon number, $m_0=\gamma_\perp^2/2g_0^2$, and the critical atom number, $N_0=2\gamma_\perp \kappa/g_0^2$, must both be much less than unity.   For conversions:
\begin{equation} \kappa=\frac{\pi c}{\lambda Q},\;\;\; \mathcal{F}=\frac{\pi c}{2L\kappa}=\frac{\lambda Q}{2L},\end{equation} where $L$ is the cavity length, $Q$ is the quality factor, and $F$ is the finesse.

The inverse of the critical atom number is known in the optical bistability literature as the cooperativity parameter $C=1/N_0$~\cite{LugiatoBook84}.  Bistability ensues when the {\it{collective cooperativity}} $NC>1$, where $N$ is the intracavity particle number.
Other authors~\cite{Vuletic00,Vuletic01, Vuletic03_near, Vuletic03_far, Vuletic04, Vuletic05} choose to describe the cooperativity not in terms of the notation of $g_0$, etc., but rather in the more experimentally physical language of scattering, which is applicable in the weak coupling regime.  This is often less convenient when working with the master equation, and the following describes how to convert between the two descriptions of the cooperativity.  The collective cooperativity may be reframed as the product of the number of intracavity photon roundtrips ($F/\pi$) and the fraction of scattered light captured by the cavity mode, $2\Delta\Omega=2\cdot3/(k^2w_0^2)$, where $k=2\pi/\lambda$ and $w_0$ is the cavity mode waist:  \begin{equation} 
C=2\Delta\Omega F/\pi=6F/(\pi k^2w^2)=\eta.
\end{equation}  This Purcell factor~\cite{Purcell}, defined as $\eta=3Q\lambda^3/4\pi^2V$ is none other than the cooperativity, and is typically the preferred notation in semiclassical treatments. The following demonstrates their equivalence:\be\label{coop}  
C=g_0^2\cdot\frac{1}{\kappa\gamma_{\perp}}=\frac{3c\lambda^2\gamma_{\perp}}{4\pi V_m}\cdot \frac{2LF}{\pi c\gamma_{\perp}}=\frac{3Q\lambda^3}{4\pi^2V}.
\ee
Since $k=2\pi/\lambda$ and $w_0^2\propto\ V_m/(L\pi)$, upon substitution we have:
\be\label{coop2}
C=\frac{6F}{\pi k^2w_0^2}=\eta.
\ee
Similarly, the collective enhancement, $N\eta$, is simply the collective cooperativity, $NC$.

\section{Candidate OH cooling transitions}\label{OH_transitions}
\begin{figure}[h]
\begin{center}
\scalebox{0.47}[0.470]{\includegraphics{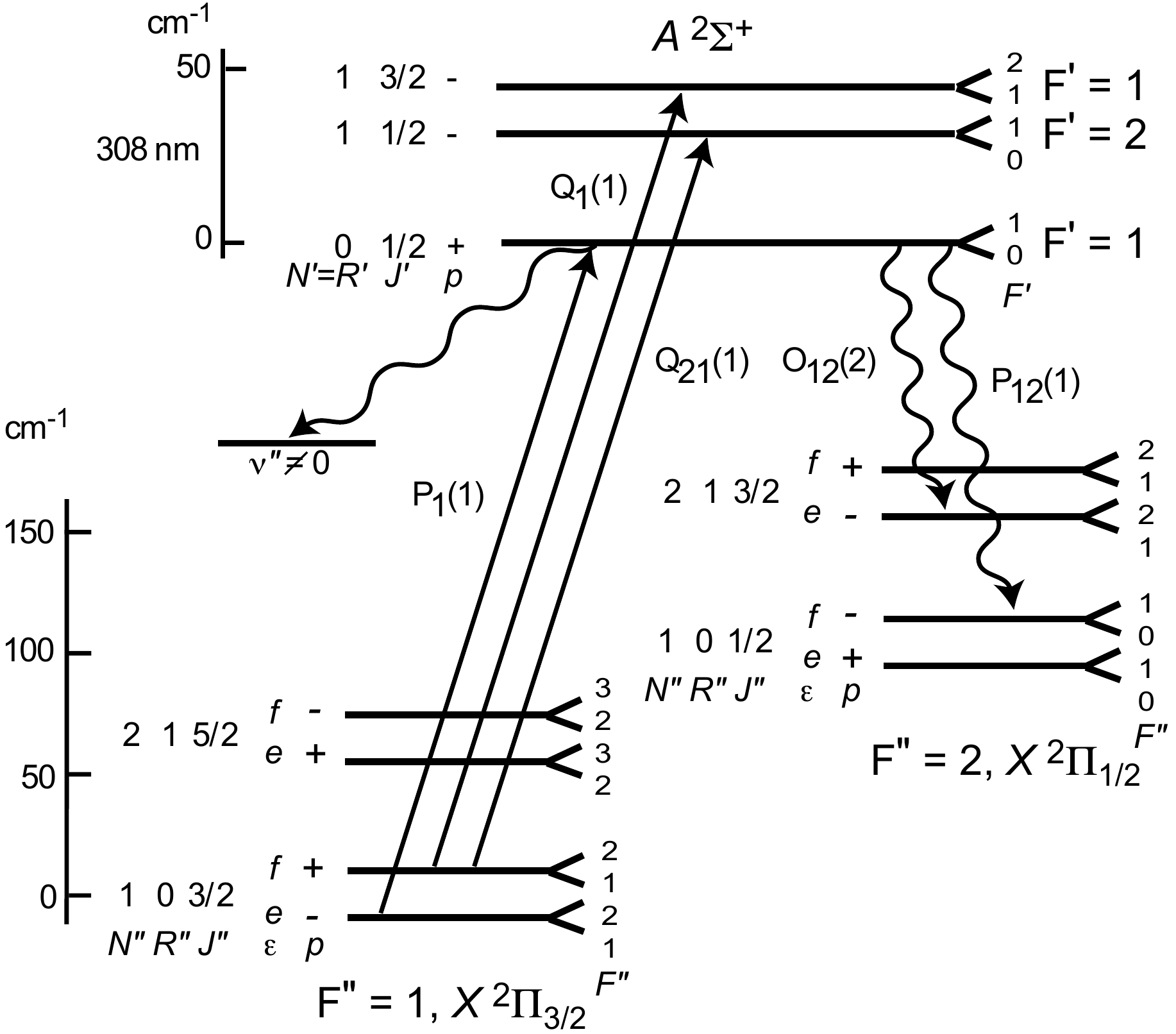}}
\caption{\label{fig:OHlevels} OH level diagram depicting the P$_1(1)$, Q$_1$(1), and Q$_{21}$(1) electronic transitions.   The first excited level $A\ ^2\Sigma^+$ and the two $X\ ^2\Pi$ grounds states are shown.  For clarity, only the P$_1(1)$ decay channels are shown: P$_{12}(1)$, O$_{12}(2)$, and the $\nu'=0\rightarrow\nu''\neq0$ transitions. }
\end{center}
\end{figure} 
The Stark decelerator provides samples of weak-field seeking F$''=1$, $X^2\Pi_{3/2}$ OH molecules~\cite{Bochinski03}.  F$''=1$ is spectroscopic notation for the lower energy state of a doublet (F$=1$, $J=N+1/2$, while for F$=2$, $J=N-1/2$) and is not the hyperfine quantum number, $F$, which is italicized.   The double-prime denotes ground state quantum numbers, while a single prime refers to the excited state. The total angular momentum is equal to $\vec{J}=\vec{L}+\vec{S}+\vec{R}$, where the three vectors are orbital, spin, and rotational angular momentum, respectively.  The angular momentum $N$ is defined as $\vec{N}=\vec{R}+\vec{L}$.

The majority of the OH ground state population arrives at the terminus of the decelerator in the $\nu''=0$, $F''=2$, $\epsilon=f$ symmetry, and $p=+$ parity state.    Microwaves can easily transfer this population to the $e,-$ lower $\Lambda$-doublet states~\cite{Hudson06,Lev06}.  The total angular momentum in the F$''=1$, $X^2\Pi_{3/2}$ ground state is $J''=3/2$.  The orbital plus rotational angular momentum quantum number in this ground state is $N''=R''+L''=1$. 

Electric dipole transitions are allowed between transitions of opposite parity that satisfy $\Delta J=0$, $\pm1$, and $\Delta N=0$, $\pm1$, $\pm2$, where $\Delta J=J'-J''$ and $\Delta N=N'-N''$. The strongest electronic cooling transitions that originate in the F$''=1$, $X^2\Pi_{3/2}$ ground state are P$_1$(1),  Q$_1$(1), and Q$_{21}$(1).  The spectroscopic notation is defined as~\cite{Dieke62}: $\Delta N_{\mbox{F}'\mbox{F}''}(N'')$, where $\Delta N\equiv[-2,-1,0,1,2]\rightarrow[\mbox{O,\,P,\,Q,\,R,\,S}]\).  In this notation, if $\mbox{F}'=\mbox{F}''$, then only one subscript index is used.

Table~\ref{OH} lists the properties of the $^2\Pi_{3/2}$ to $^2\Sigma^+_{1/2}$ ($\Delta v=0$ band) transitions of interest.  To cool on the P$_1(1)$ transition, a single microwave pulse is needed to drive the population to the $e$ state.  The other two electronic transitions need either one additional microwave pulse (Q$_1(1)$) or no microwaves at all (Q$_{21}(1)$) for initial state preparation.  Only Q$_1(1)$ offers the possibility of a $\Delta F=+1$ transition, which allows the molecule to easily return to the same Zeeman hyperfine ground state after each Rayleigh scattered photon.  The recoil frequency for these transitions is: $\omega_{rec}=2\pi\cdot1.23\times10^5$ Hz.  

With two repumpers, one each on the P$_{12}(1)$ and the O$_{12}(2)$ lines, the P$_1(1)$ transition remains 0.4\% open due to the possibility to scatter to higher vibrational levels (i.e., $\nu\neq0$).  Closing the Q$_1(1)$ transition to 4\% requires---in order of importance---repumpers on the P$_{12}(2)$, Q$_{12}(1)$, P$_1(2)$, Q$_{12}(3)$ lines.  To close the Q$_{21}(1)$ transition to 1\% requires repumping the Q$_2(1)$ and P$_2(2)$ lines. Figure~\ref{fig:OHlevels} sketches the relevant energy levels for the electronic transitions, but for clarity only includes the decay channels from the P$_1(1)$ transition. The first vibrational transition, Q$\frac{3}{2}e$, may be completely closed with five repumpers and is $\sim$60\% closed with just one repumper on the P$\frac{5}{2}f$ line, where the notation is $\Delta J(J'')\epsilon$ and $\epsilon=e,f$.

\end{document}